\documentclass[a4paper,11pt]{article}
\pdfoutput=1
\usepackage{jheppub}
\usepackage[T1]{fontenc}
\usepackage{bm,amsmath,amssymb,color,bbm,mathrsfs,slashed,xspace} 
\usepackage[shortlabels]{enumitem}

\DeclareMathOperator\rme{\mathrm{e}}
\renewcommand{\Im}{\mathrm{Im}}
\renewcommand{\Re}{\mathrm{Re}}
\newcommand{\der}{\partial}
\renewcommand{\bar}[1]{\overline{#1}}

\newcommand{\bep}{\begin{pmatrix}} 
\newcommand{\eep}{\end{pmatrix}}

\newcommand{\U}{\text{U}}

\newcommand{\CC}{\mathbb{C}}

\renewcommand{\epsilon}{\varepsilon}
\newcommand{\rmd}{\mathrm{d}}

\def\wh#1{\widehat{#1}}
\newcommand{\pp}{\mathbf{p}}
\newcommand{\muu}{\wh{\mu}}
\newcommand{\csb}{\text{$\chi$SB}\xspace}

\def\ba#1\ea{\begin{align}#1\end{align}}
\def\mkakko#1{\left( #1 \right)}
\def\ckakko#1{\left\{ #1 \right\}}
\def\kkakko#1{\left[ #1 \right]}

\title{Non-Hermitian BCS-BEC crossover of Dirac fermions}

\author[a]{Takuya Kanazawa}

\affiliation[a]{Research and Development Group, Hitachi, Ltd., Kokubunji, Tokyo 185-8601, Japan}
\emailAdd{tkanazawa@nt.phys.s.u-tokyo.ac.jp}

\abstract{We investigate chiral symmetry breaking in a model of Dirac fermions with a complexified coupling constant whose imaginary part represents dissipation. We introduce a chiral chemical potential and observe that for real coupling a relativistic BCS-BEC crossover is realized. We solve the model in the mean-field approximation and construct the phase diagram as a function of the complex coupling. It is found that the dynamical mass increases under dissipation, although the chiral symmetry gets restored if dissipation exceeds a threshold.}

\allowdisplaybreaks 
\begin{document} 
\maketitle
\flushbottom

\section{Introduction}
One of the fundamental tenets of modern quantum mechanics and quantum field theory is that  Hamiltonians should be Hermitian. It ensures that energy levels are real and the time evolution is unitary. However, it has been perceived in recent years that there are situations in which physics can be effectively described in terms of non-Hermitian Hamiltonians \cite{Moiseyev2011,Ashida:2020dkc}. Such a description proves to be useful for open quantum systems that interact with environments \cite{Dalibard:1992zz,Carmichael_1993,Daley_2014}. In $\mathcal{PT}$-symmetric quantum mechanics, the existence of $\mathcal{PT}$ symmetry ensures real energy spectra even when the Hamiltonian is not Hermitian \cite{Bender:1998ke,Bender:2002vv,Bender:2007nj}. This is not just a mathematical possibility but can be realized experimentally \cite{Takasu2020pt}. It is not an overstatement to say that the physics of non-Hermitian Hamiltonians is far richer than conventional Hermitian ones; in fact, symmetry classification \cite{bernard2001classification,Kawabata:2018gjv} reveals that there are 38 distinct symmetry classes in non-Hermitian systems, while there are only 10 in Hermitian systems. There are a plethora of exotic phenomena that are unique to non-Hermitian systems, such as the non-Hermitian skin effect \cite{Martinez_Alvarez_PRB_2018,Yao_2018,Martinez_Alvarez_EPJ_2018,Borgnia:2020mkg,Okuma_2020} and the non-Hermitian localization transitions \cite{Hatano_1996}. 

Non-Hermitian formulations can nowadays be found not only in condensed matter physics but also in high-energy physics. In studies of hot QCD, it is one of standard approaches to quarkonia to postulate a complex-valued interquark potential \cite{Rothkopf:2011db,Rothkopf:2019ipj,Akamatsu:2020ypb}; the non-Hermitian chiral magnetic effect has been proposed \cite{Chernodub:2019ggz}; relativistic quantum field theories with a complex-valued coupling have been investigated \cite{Denbleyker:2007dy,Denbleyker:2010sv,Meurice:2011nf,Liu:2011zzh,Denbleyker:2011aa,Bazavov:2012ex}. There are also works on non-Hermitian deformations of the Nambu--Jona-Lasinio (NJL) model \cite{Bender:2005hf,Alexandre:2015kra,Alexandre:2017foi,Beygi:2019qab,Alexandre:2020wki,Alexandre:2020bet,Felski:2020vrm,Alexandre:2020tba,Mavromatos:2020hfy,Chernodub:2020cew}, which is an archetypal model for dynamical chiral symmetry breaking (\csb) in QCD \cite{Nambu:1961tp,Nambu:1961fr,Klevansky:1992qe,Hatsuda:1994pi,Buballa:2003qv}. It is also worthwhile to mention that the research of a non-Hermitian Dirac operator has a long history in QCD --- in the presence of quark chemical potential, the Euclidean Dirac operator is no longer anti-Hermitian and its eigenvalues spread over the complex plane. A complex action in the path integral then leads to a serious sign problem in numerical simulations on a lattice and a number of cures have been proposed in the literature \cite{Guralnik:2007rx,Alexanian:2008kd,Witten:2010cx,Cristoforetti:2012su,Fujii:2013sra,Tanizaki:2014xba,Kanazawa:2014qma,Tanizaki:2015rda,Fujii:2015vha,Alexandru:2015sua,Alexandru:2015xva,Mori:2017pne,Alexandru:2017czx,Fukuma:2019wbv,Mou:2019gyl} (see \cite{Alexandru:2020wrj} for a recent review). Studies of complex spectra of the Dirac operator have revealed a deep link between QCD at finite density and random matrix theory for non-Hermitian quantum chaos \cite{Markum:1999yr,Khoruzhenko2009,Kanazawabook}. 

Superfluidity and superconductivity are amongst the most salient and fascinating phenomena in quantum many-body physics \cite{LeggettBook2006,Casalbuoni:2018haw}. Non-Hermitian fermionic superfluidity has been explored in  \cite{Chtchelkatchev2012,Ghatak2018,Zhou2018,UedaPRL2019,Iskin2020}.%
\footnote{Properties of non-equilibrium fermionic systems with attractive interactions (such as the exciton BEC \cite{Blatt1962,Yoshioka_2011}) were theoretically studied in e.g.  \cite{Ogawa2007,Yamaguchi2012,Hanai_JLTP_2016,Hanai_PRB_2017,Kawamura_JLTP_2019,Kawamura_PRA_2020}. Excitonic states in optically-pumped Dirac materials were investigated in \cite{Triola_2017,Pertsova_2018,Pertsova_review2020}.}
In \cite{UedaPRL2019} the authors pointed out that ultracold atomic gases with two-body losses due to inelastic collisions can be naturally described with a complex-valued interaction. They solved the gap equation for fermions on a square/cubic lattice and mapped out the phase diagram in the mean-field approximation. In \cite{Iskin2020} this analysis was extended to fermions in a continuum model, where the phase diagram across the entire range from weak to strong coupling was obtained.  This is a non-Hermitian analog of the well-established BCS-BEC crossover of fermions with $s$-wave interactions \cite{Nozieres1985,Chen2005,LeggettBook2006,GiorginiRMP2008,ZwergerBook2012,RanderiaBook2014,Strinati2018} --- when the $s$-wave scattering length is varied, the system evolves continuously from a weakly interacting BCS regime of loose Cooper pairs to the BEC regime of tightly bound molecules. 

The primary goal of the present paper is to generalize the analysis of \cite{UedaPRL2019,Iskin2020} to Dirac fermions and investigate a non-Hermitian relativistic BCS-BEC crossover.%
\footnote{Although \cite{Kanazawa:2014qma} solved a zero-dimensional model of Dirac fermions with a complex four-fermion coupling, its higher-dimensional analog has not been thoroughly studied yet.} What is the motivation of this study? First, Dirac fermions can be realized with ultracold atoms loaded on an optical lattice \cite{Zhu_2007,Lim_2009,IgnacioCirac:2010us} and hence the experimental protocol proposed in \cite{UedaPRL2019} to produce a complex coupling can in principle be applied to this case as well. Second, a relativistic BCS-BEC crossover is believed to take place in QCD-like theories at finite density \cite{Nishida:2005ds,Abuki:2006dv,Deng:2006ed,He:2007kd,He:2007yj,Sun:2007fc,Brauner:2008td,Chatterjee:2008dr,He:2010nb,Kanazawa:2011tt} (see \cite{He:2013gga} for a review). While the interaction strength and the density can be separately varied in nonrelativistic systems, this is not the case in gauge theories --- actually we face a density-induced crossover: at low density, quark matter is strongly coupled and tightly bound diquarks condense, whereas at high density the coupling is weak due to asymptotic freedom and a BCS-type description is justified \cite{Alford:2007xm}. The study of such a crossover (including the possibility of phase transitions at intermediate densities) is potentially relevant to compact star phenomenology and heavy-ion collision experiments. 

In this paper we investigate \csb in the NJL model with a complex coupling. Generally, \csb is considered to be a strong coupling phenomenon and a weakly coupled BCS picture does not apply. However, at finite chiral chemical potential \cite{Fukushima:2008xe,Yamamoto:2011gk,Ruggieri:2011xc,Gatto:2011wc,Braguta:2015zta,Xu:2015vna,Braguta:2016aov,Chernodub:2019ggz,Braguta:2019pxt}, \csb occurs at an arbitrarily weak coupling due to the fact that a chiral chemical potential induces a nonzero density of states for fermions at low energy and serves as a catalyst of \csb \cite{Braguta:2016aov,Braguta:2019pxt}. By tuning both the coupling strength and the chiral chemical potential, we probe the entire range of the BCS-BEC crossover for \csb and construct a complete phase diagram as a function of the complex four-fermion coupling. A novel mechanism for emergence and disappearance of complex saddles of the action is also illustrated. 

Two caveats are in order here. First, to keep the presentation as simple as possible, we will not consider superfluidity (diquark condensation) in this work. Second, although a chirally imbalanced matter in gauge theories is intrinsically unstable due to the axial anomaly \cite{Redlich:1984md,Rubakov:1985nk,Akamatsu:2013pjd}, our model has no coupling to gauge fields and there is no instability due to anomalies. 

This paper is organized as follows. In section~\ref{sc:def} the model is defined and the thermodynamic potential is derived. In section~\ref{sc:real} the phase diagram for real coupling is presented. The detrimental effect of the baryon chemical potential on \csb is illustrated. In section~\ref{sc:complex} we turn on an imaginary part of the coupling, solve the gap equation numerically, map out the phase diagram, and determine the boundary between the normal phase, a metastable \csb phase, and a stable \csb phase. The results are then compared with those for nonrelativistic fermions \cite{UedaPRL2019,Iskin2020}. In section~\ref{app} the dependence on the UV regularization scheme is discussed. We conclude in section~\ref{sc:conc}.

\section{\label{sc:def}The NJL model}

We consider a model with the partition function $\displaystyle Z=\int \mathcal{D}\bar\psi \mathcal{D}\psi~\exp\mkakko{-\int_0^\beta \rmd \tau \int \rmd^3x ~\mathcal{L}}$ where the Euclidean Lagrangian is given by
\ba
	\mathcal{L} & = \bar\psi(\slashed{\der}
	-\mu \gamma_0-\mu_5\gamma_5\gamma_0)\psi
	- \frac{G}{2}\kkakko{(\bar\psi\psi)^2+(\bar\psi i\gamma_5 \psi)^2}. 
	\label{eq:24543543}
\ea
It is invariant under $\U(1)_V\times \U(1)_A$ symmetry transformations. $\mu$ is the quark chemical potential and $\mu_5$ is the chiral (or axial) chemical potential. We set the current mass to zero. Eq.~\eqref{eq:24543543} is the same model as in \cite{Braguta:2016aov,Braguta:2019pxt} except that here we have nonzero $\mu$. It is straightforward to introduce $N_f$ flavors of quarks and let $N_f\to \infty$ with $G\propto 1/N_f$, which allows us to rigorously justify a saddle point analysis, but we shall stick to $N_f=1$ for simplicity of exposition.   

With the Hubbard-Stratonovich transformation, quarks can be readily integrated out and yields
\ba
	Z & = \int \mathcal{D}\sigma\mathcal{D}\pi~
	\rme^{-\frac{1}{2G}\int (\sigma^2+\pi^2)} 
	\det\mkakko{\slashed{\der}-\mu\gamma_0-\mu_5\gamma_5\gamma_0-\sigma-i\gamma_5\pi} .
\ea
The bosonic fields are related to fermionic observables as $\langle \sigma \rangle=G\langle \bar\psi\psi \rangle$ and $\langle \pi \rangle=G\langle \bar\psi i\gamma_5 \psi \rangle$. In the mean-field approximation, we have
\ba
	Z & = \int \rmd \sigma \int \rmd \pi ~ 
	\rme^{-\frac{V_4}{2G}(\sigma^2+\pi^2)}
	\prod_{p_0}\prod_{\pp}
	\kkakko{(p_0+i\mu)^2+E_+^2}\kkakko{(p_0+i\mu)^2+E_-^2}
	\\
	& = \int \rmd \sigma \int \rmd \pi ~ 
	\rme^{-\frac{V_4}{2G}(\sigma^2+\pi^2)}\Bigg\{
	\prod_{p_0}\prod_{\pp}\kkakko{p_0^2+(E_++\mu)^2}
	\kkakko{p_0^2+(E_+-\mu)^2}
	\notag
	\\
	& \quad \times \kkakko{p_0^2+(E_-+\mu)^2}
	\kkakko{p_0^2+(E_--\mu)^2}
	\Bigg\}^{1/2}
	\hspace{-10pt}
	\\
	& \propto \int \rmd \sigma \int \rmd \pi ~ 
	\rme^{-\frac{V_4}{2G}(\sigma^2+\pi^2)}\prod_{\pp}
	\bigg[\cosh\mkakko{\frac{E_++\mu}{2T}}
	\cosh\mkakko{\frac{E_+-\mu}{2T}}
	\notag
	\\
	& \quad \times 
	\cosh\mkakko{\frac{E_-+\mu}{2T}}
	\cosh\mkakko{\frac{E_--\mu}{2T}}\bigg]\,,
\ea
where $V_4\equiv V_3/T$ is the Euclidean spacetime volume, 
\ba
	E_{\pm} & \equiv \sqrt{(|\pp|\pm \mu_5)^2+\sigma^2+\pi^2}, 
	\label{eq:E243589}
\ea
and we used \cite{dlmf_cosh} $\displaystyle \prod_{n=-\infty}^{\infty}\kkakko{1+\frac{z^2}{(2n+1)^2\pi^2}}=\cosh^2\mkakko{\frac{z}{2}}$. Then $\displaystyle Z = \int \rmd \sigma \int\rmd \pi~ \rme^{-V_4 \mathcal{S}}$ with
\ba
	\mathcal{S} & \equiv 
	\frac{\sigma^2+\pi^2}{2G} - \frac{T}{2\pi^2}\int_0^\Lambda \rmd p\; p^2 
	\bigg[
		\log \cosh\mkakko{\frac{E_++\mu}{2T}} 
		+ \log \cosh\mkakko{\frac{E_+-\mu}{2T}}
	\notag
	\\& \quad 
		+ \log \cosh\mkakko{\frac{E_-+\mu}{2T}}
		+ \log \cosh\mkakko{\frac{E_--\mu}{2T}}
	\bigg]\,,
	\notag
\ea
where a momentum cutoff $\Lambda$ was introduced to remove UV divergences. (See section~\ref{app} for a discussion on an alternative UV regularization.) Let us define dimensionless variables
\ba
	g \equiv G\Lambda^2, \quad t \equiv \frac{T}{\Lambda}, \quad 
	M^2 \equiv \frac{\sigma^2+\pi^2}{\Lambda^2}, \quad \muu \equiv \frac{\mu}{\Lambda}, 
	\quad \muu_5 \equiv \frac{\mu_5}{\Lambda}\,,
\ea
which leads to a dimensionless action $S \equiv \mathcal{S}/\Lambda^4$ given by
\ba
	S & = \frac{M^2}{2g} - \frac{t}{2\pi^2}\int_0^1 \rmd x\; x^2 \Bigg[
		\log \cosh \mkakko{\frac{\sqrt{(x+\muu_5)^2+M^2}+\muu}{2t}}
		+ \log \cosh \mkakko{\frac{\sqrt{(x+\muu_5)^2+M^2}-\muu}{2t}}
	\notag
	\\
	& \quad + \log \cosh \mkakko{\frac{\sqrt{(x-\muu_5)^2+M^2}+\muu}{2t}}
	+ \log \cosh \mkakko{\frac{\sqrt{(x-\muu_5)^2+M^2}-\muu}{2t}}
	\Bigg]\,.
	\label{eq:8564323}
\ea
In the following, we will assume $0<\muu_5<1$ so that the Fermi surface stays inside the domain of integration. In generic open quantum systems, temperature is not well defined, and we treat $t$ as a formal parameter used to define the path integral for the partition function, as in  \cite{UedaPRL2019}. We will set $t$ to zero in the ensuing analysis.

\section{\label{sc:real}Phase diagram for real coupling}

\begin{figure}[tb]
	\centering
	\includegraphics[width=.45\textwidth]{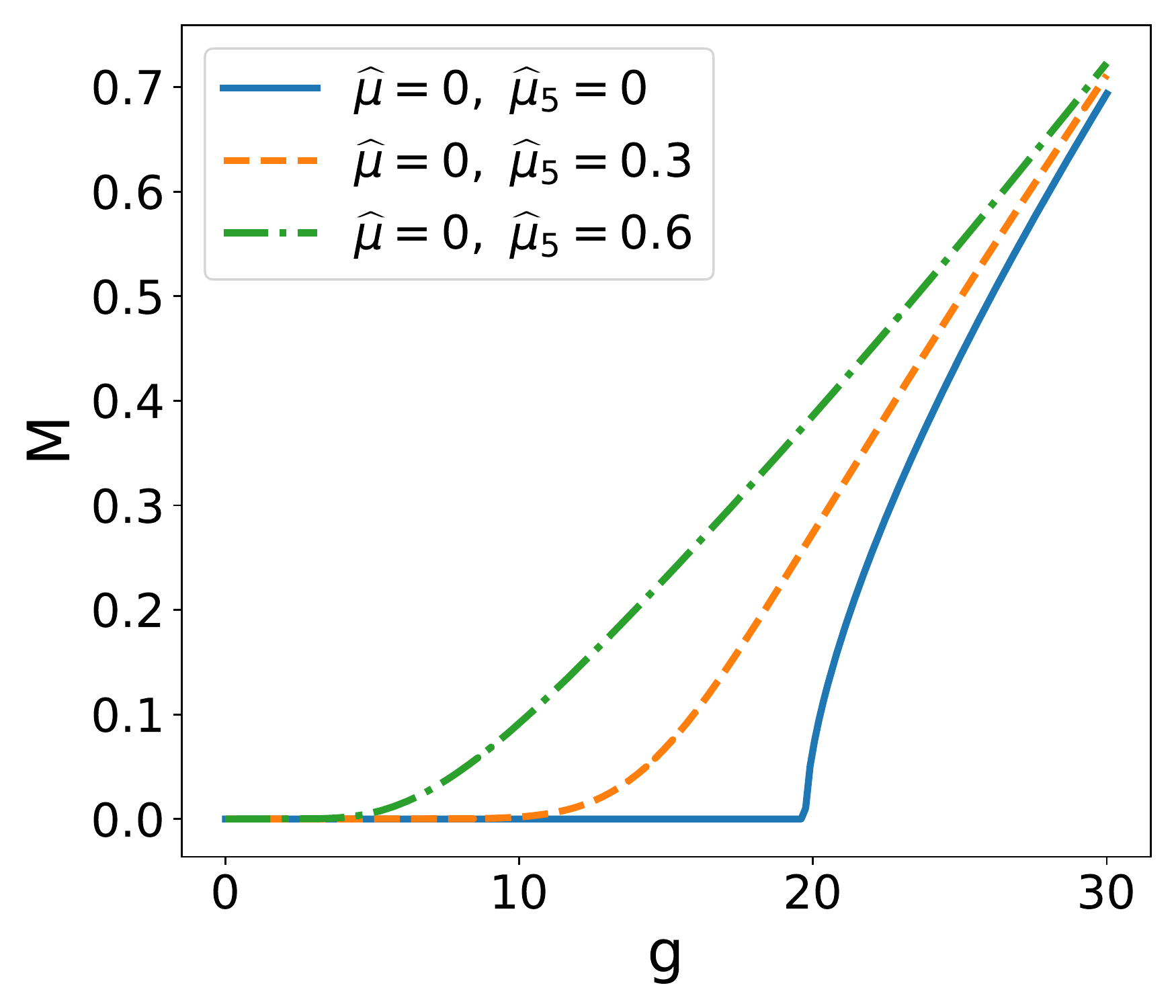}
	\ \ 
	\includegraphics[width=.5\textwidth]{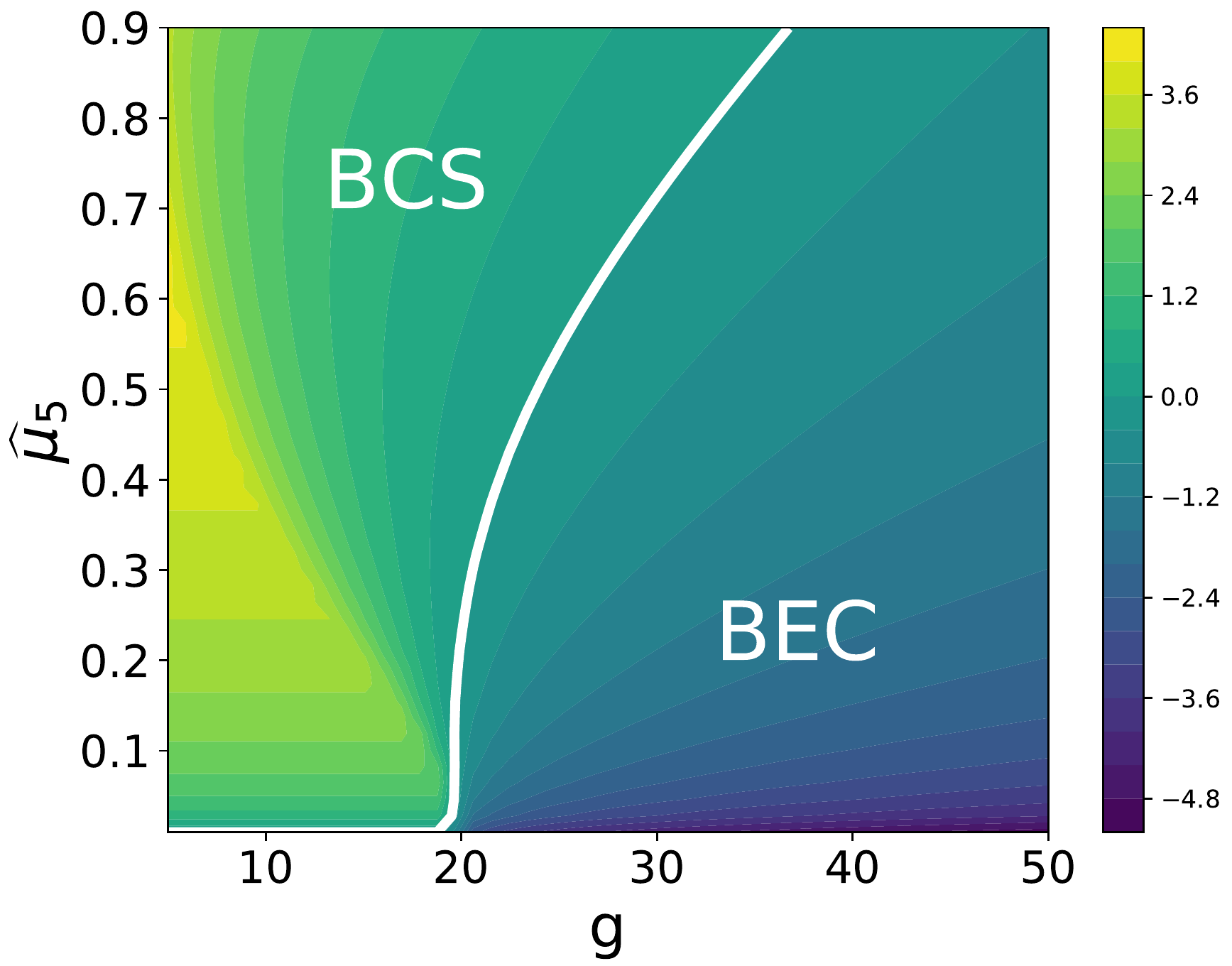}
	\caption{\label{fg:3984675}Dynamical mass (left) and the phase diagram (right) for $\mu=0$ at zero temperature. In the right plot, the color scale represents $\log(\muu_5/M)$. The white curve indicates $\log(\muu_5/M)=0$.}
\end{figure}

Let us begin with a discussion for real coupling $g>0$. In the zero-temperature limit $t\to +0$, \eqref{eq:8564323} reduces to
\ba
	S & = \frac{M^2}{2g} - \frac{1}{4\pi^2}\int_0^1 \rmd x\; x^2 \Big(
		|\sqrt{(x+\muu_5)^2+M^2}+\muu| + |\sqrt{(x+\muu_5)^2+M^2}-\muu|
	\notag
	\\
	& \quad + |\sqrt{(x-\muu_5)^2+M^2}+\muu | + |\sqrt{(x-\muu_5)^2+M^2}-\muu |
	\Big)\,.
\ea
Numerical minimization of $S$ allows us to determine the dynamical mass $M$ as a function of $g$ and $\muu_5$. Our result for $\muu=0$ is presented in Figure~\ref{fg:3984675}. As one can see from the left plot, while there is a nonzero critical coupling $\approx 20$ at $\muu_5=0$, it goes away for $\muu_5>0$: \csb occurs for any nonzero coupling. This is the catalysis effect emphasized in e.g., \cite{Braguta:2016aov,Braguta:2019pxt}. The dynamical mass increases monotonically with $\muu_5$.

What is the distinction between the BCS regime and the BEC regime? In nonrelativistic systems the unitarity limit marks a crisp boundary, but what about relativistic systems? The locus of a ``transition'' between the two regimes is not uniquely defined. As suggested in \cite{He:2013gga}, one popular criterion is to see the dispersion relation $E(\pp)$ of quasiparticles. If $E(\pp)$ takes a minimum at $|\pp|\approx p_\text{F}$, it is in the BCS regime, and if $E(\pp)$ is a monotonically increasing function of $|\pp|$, it is in the BEC regime. This works pretty well in dense QCD where $E(\pp)=\sqrt{(\sqrt{\pp^2+M^2}-\mu)^2+\Delta^2}$ (with $\Delta$ the superconducting gap) experiences such a transition when the dynamical mass $M$ is equal to $\mu$. In the current model, however, the dispersion \eqref{eq:E243589} takes a minimum at $|\pp|=\mu_5$ regardless of the dynamical mass. Yet another criterion is to compare the interparticle distance and the size of Cooper pairs. If the wave function of Cooper pairs extends beyond the average interparticle distance, it is in the BCS regime, otherwise in the BEC regime. In \cite{Abuki:2001be,Itakura:2002vr} the size of Cooper pairs in relativistic color superconductors was computed as a function of the quark density and such a BCS-BEC-type crossover was indicated. Unfortunately, a similar analysis is difficult for our model because the interaction is pointlike and the gap has no momentum dependence. As a rule of thumb, let us take the inverse of $M$ as the size of a Cooper pair, and take the inverse of $\mu_5$ as the average inter-quark distance. Then the region with $1/M>1/\muu_5$ ($1/M<1/\muu_5$) corresponds to the BCS (BEC) regime, respectively. According to this crude estimate we labeled each regime in the right plot of Figure~\ref{fg:3984675}.

\begin{figure}[tb]
	\centering
	\includegraphics[width=.47\textwidth]{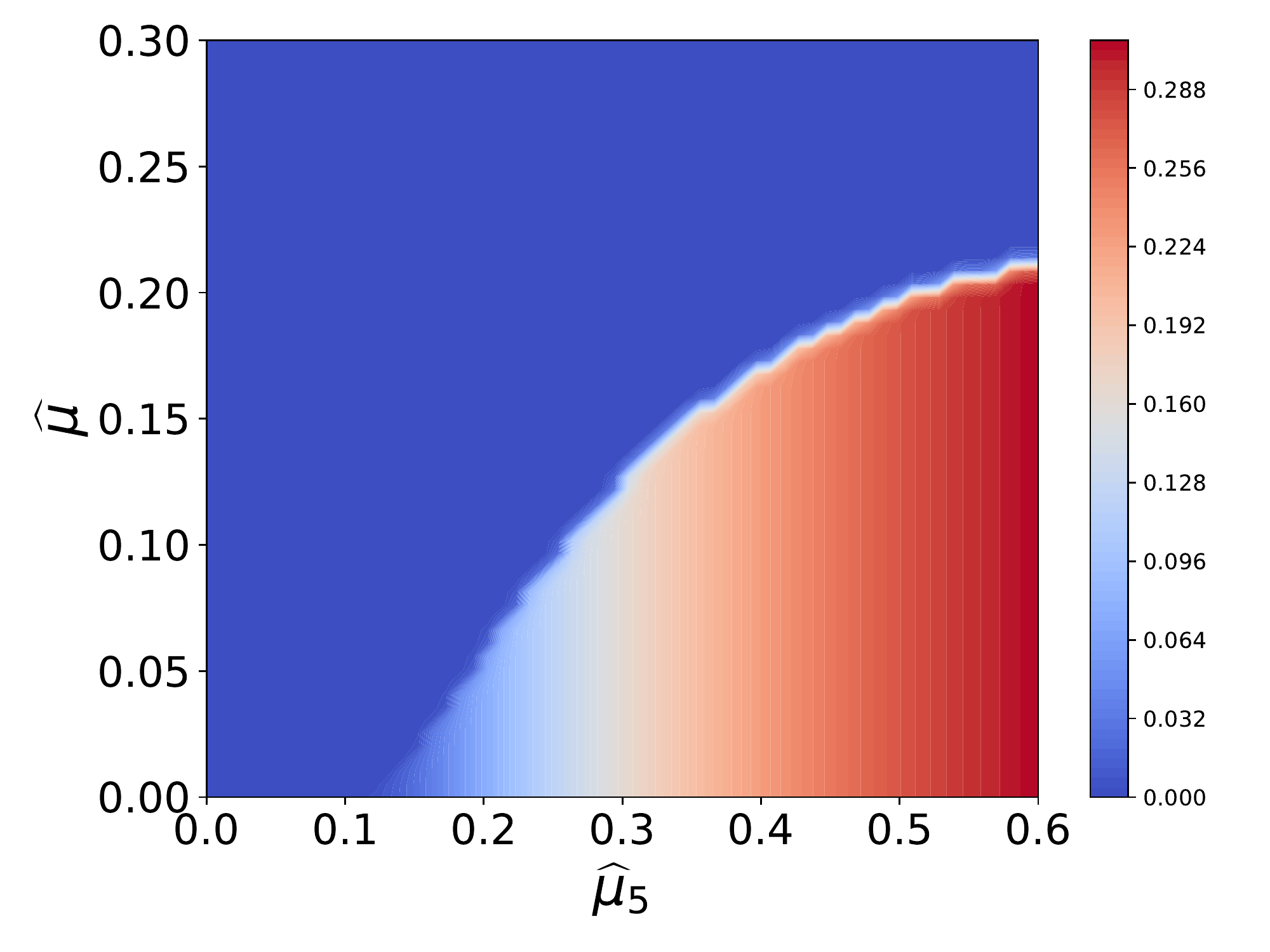}
	\ \ 
	\includegraphics[width=.47\textwidth]{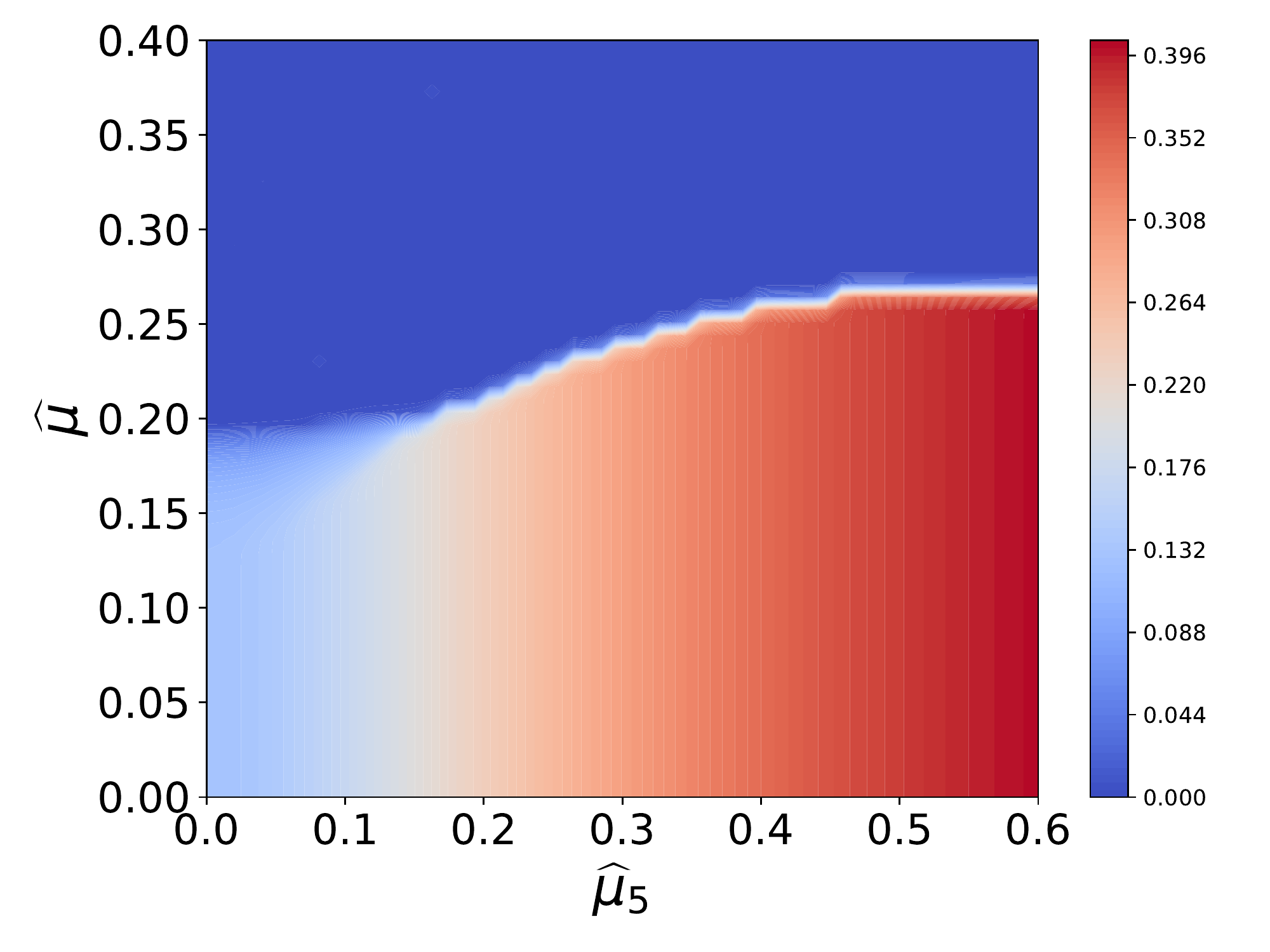}
	\caption{\label{fg:stress}Phase diagrams for $g=17.5$ (left) and $g=20.5$ (right) at zero temperature. The color scale represents $M$.}
\end{figure}

Next we proceed to the analysis for nonzero $\muu$. The phase diagrams are shown in Figure~\ref{fg:stress}. The main observation here is that for any $\muu_5$, there is a critical $\muu$ beyond which the chiral symmetry is restored. This is because $\muu$ induces a mismatch of Fermi surfaces and disrupts Cooper pairing. The critical value of $\muu$ is known as the Chandrasekhar-Clogston limit \cite{Chandrasekhar1962,Clogston1962}. Analogous situations arise in both condensed matter \cite{Radzihovsky_2010,Chevy_2010} and high-density QCD \cite{Alford:2000ze,Splittorff:2000mm,Bedaque:2001je,Casalbuoni:2003wh,Nickel:2009wj,Buballa:2014tba}. When the ordinary isotropic Cooper pairing is hampered, a nonstandard pairing that breaks translation symmetry is likely to set in, though it is beyond the scope of this paper.

\section{\label{sc:complex}Phase diagram for complex coupling}

Finally we complexify the coupling constant. Throughout this section we set $\muu=0$ to simplify the ensuing numerical analysis. Eq.~\eqref{eq:8564323} reduces to
\ba
	S & = \frac{M^2}{2g} - \frac{t}{\pi^2}\int_0^1 \rmd x\; x^2 \Bigg[
		\log \cosh \mkakko{\frac{\sqrt{(x+\muu_5)^2+M^2}}{2t}}
		+ \log \cosh \mkakko{\frac{\sqrt{(x-\muu_5)^2+M^2}}{2t}}
	\Bigg]
\ea
where $g\in\CC$ and $M\in\CC$. When $M$ is purely imaginary, $\sqrt{(x\pm \muu_5)^2+M^2}$ may be exactly on the branch cut of the complex square root, which makes the zero-temperature limit $t\to +0$ ill-defined. When $M$ is \emph{not} purely imaginary, we have for $t\to+0$
\ba
	S & =\frac{M^2}{2g} - \frac{1}{2\pi^2}\int_0^1 \rmd x\; x^2 \big[
		\sqrt{(x+\muu_5)^2+M^2} + \sqrt{(x-\muu_5)^2+M^2}
	\big] \,.
\ea
This integral can be performed with the formula
\ba
	\int \rmd x\; x^2 \sqrt{(x+\mu)^2+M^2} & = 
	- \frac{M^2(M^2-4\mu^2)}{8}\tanh^{-1}\mkakko{\frac{x+\mu}{\sqrt{(x+\mu)^2+M^2}}}
	\notag
	\\
	& \quad + \frac{\sqrt{(x+\mu)^2+M^2}}{24} \kkakko{6x^3+2\mu x^2+(3M^2-2\mu^2)x+2\mu^3-13M^2\mu}
	\notag
	\\
	& =: F(x,\mu,M)\,,
\ea
where $\tanh^{-1}(z)=\frac{1}{2}\log\mkakko{\frac{1+z}{1-z}}$ is the inverse function of $\tanh(z)$. This way we obtain
\ba
	S & = \frac{M^2}{2g} - \frac{1}{2\pi^2}\kkakko{
		F(1,\muu_5,M) - F(0,\muu_5,M) + F(1,-\muu_5,M) - F(0,-\muu_5,M)
	}
	\\
	& = \frac{M^2}{2g} - \frac{1}{2\pi^2}\Bigg[
	- \frac{M^2(M^2-4\muu_5^2)}{8}
	\Bigg\{
		\tanh^{-1}\mkakko{\frac{1+\muu_5}{\sqrt{(1+\muu_5)^2+M^2}}} 
	\notag
	\\
	& \quad + 
		\tanh^{-1}\mkakko{\frac{1-\muu_5}{\sqrt{(1-\muu_5)^2+M^2}}}
	\Bigg\}
	\notag
	\\
	& \quad 
	+ \frac{\sqrt{(1+\muu_5)^2+M^2}}{24} (6+2\muu_5+3M^2-2\muu_5^2+2\muu_5^3-13M^2\muu_5)
	\notag
	\\
	& \quad 
	+ \frac{\sqrt{(1-\muu_5)^2+M^2}}{24} (6-2\muu_5+3M^2-2\muu_5^2-2\muu_5^3+13M^2\muu_5)
	\Bigg]
	\,.
\ea
Since $S$ is a function of $M^2$, the trivial vacuum $M=0$ is always a solution to $\der S/\der M=0$. We are interested in the gap equation for $M\ne 0$, which reads
\ba
	0 & \overset{!}{=} \frac{1}{2M} \frac{\der S}{\der M} 
	\notag
	\\
	& = \frac{1}{2g} + \frac{1}{8\pi^2}\Bigg[
		(3\muu_5-1)\sqrt{(1+\muu_5)^2+M^2} - (3\muu_5+1)\sqrt{(1-\muu_5)^2+M^2}
	\notag
	\\
	& \quad 
	+ (M^2 - 2\muu_5^2) \ckakko{
			\tanh^{-1}\mkakko{\frac{1+\muu_5}{\sqrt{(1+\muu_5)^2+M^2}}}
			+ \tanh^{-1}\mkakko{\frac{1-\muu_5}{\sqrt{(1-\muu_5)^2+M^2}}}
		}
	\Bigg] \,.
	\label{eq:hgdf8}
\ea
We have varied $g$ on the complex plane and numerically searched for a solution to \eqref{eq:hgdf8} for each $g$. It turned out that there was no solution for $\Re\; g<0$, indicating that chiral symmetry is unbroken for $\Re\; g<0$. This is natural because $\Re\; g<0$ is a repulsive interaction. Furthermore, the phase structures for $\Im\;g>0$ and $\Im\;g<0$ are symmetric about the real axis of $g$. Hence we will assume $\Re\;g>0$ and $\Im\;g>0$ in the following.  

\begin{figure}[tb]
	\centering
	\includegraphics[width=.47\textwidth]{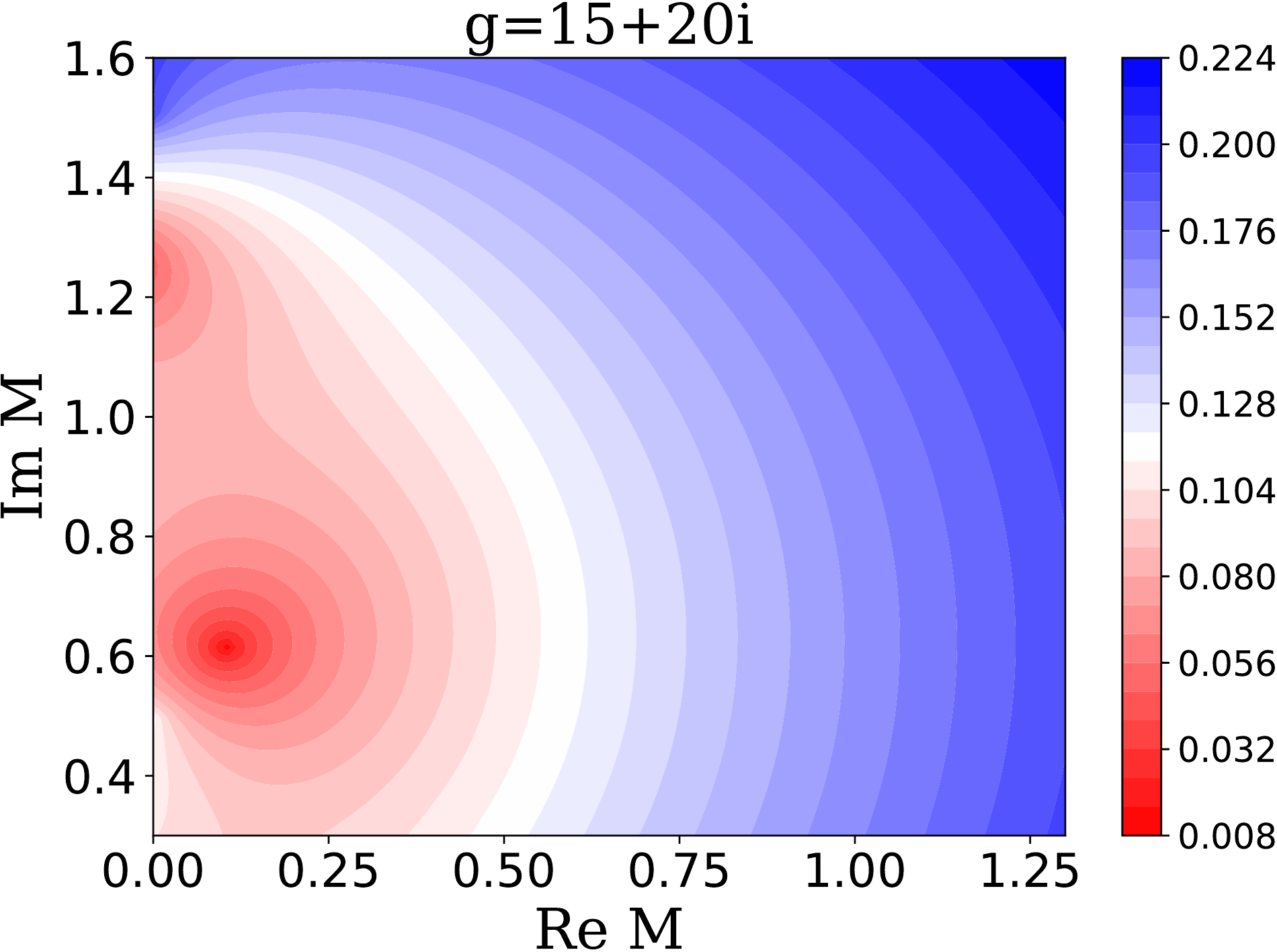} \quad 
	\includegraphics[width=.47\textwidth]{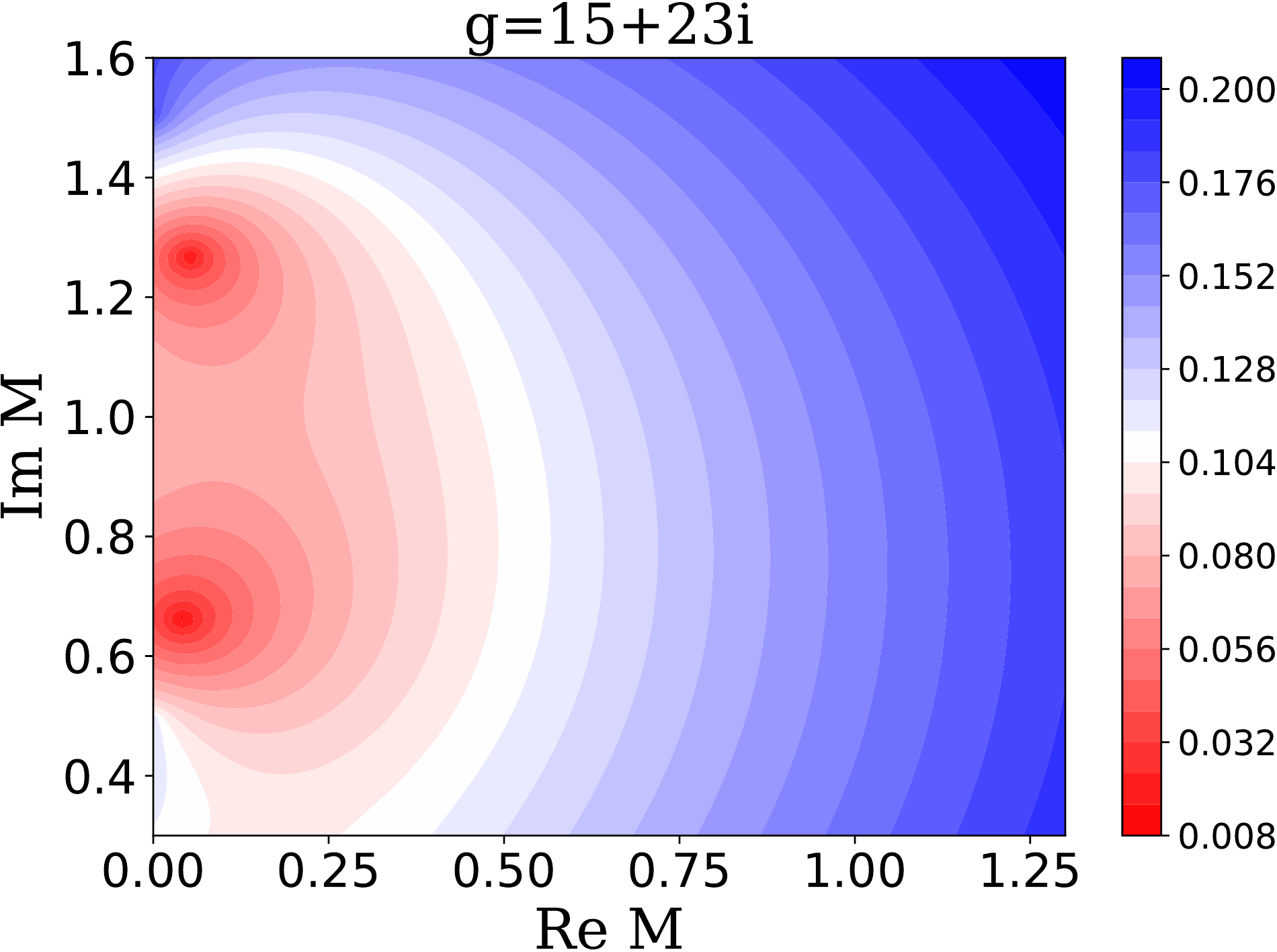}
	\put(-428,145){\Large (a)}
	\put(-212,145){\Large (b)}
	\vspace{12pt}\\
	\includegraphics[width=.47\textwidth]{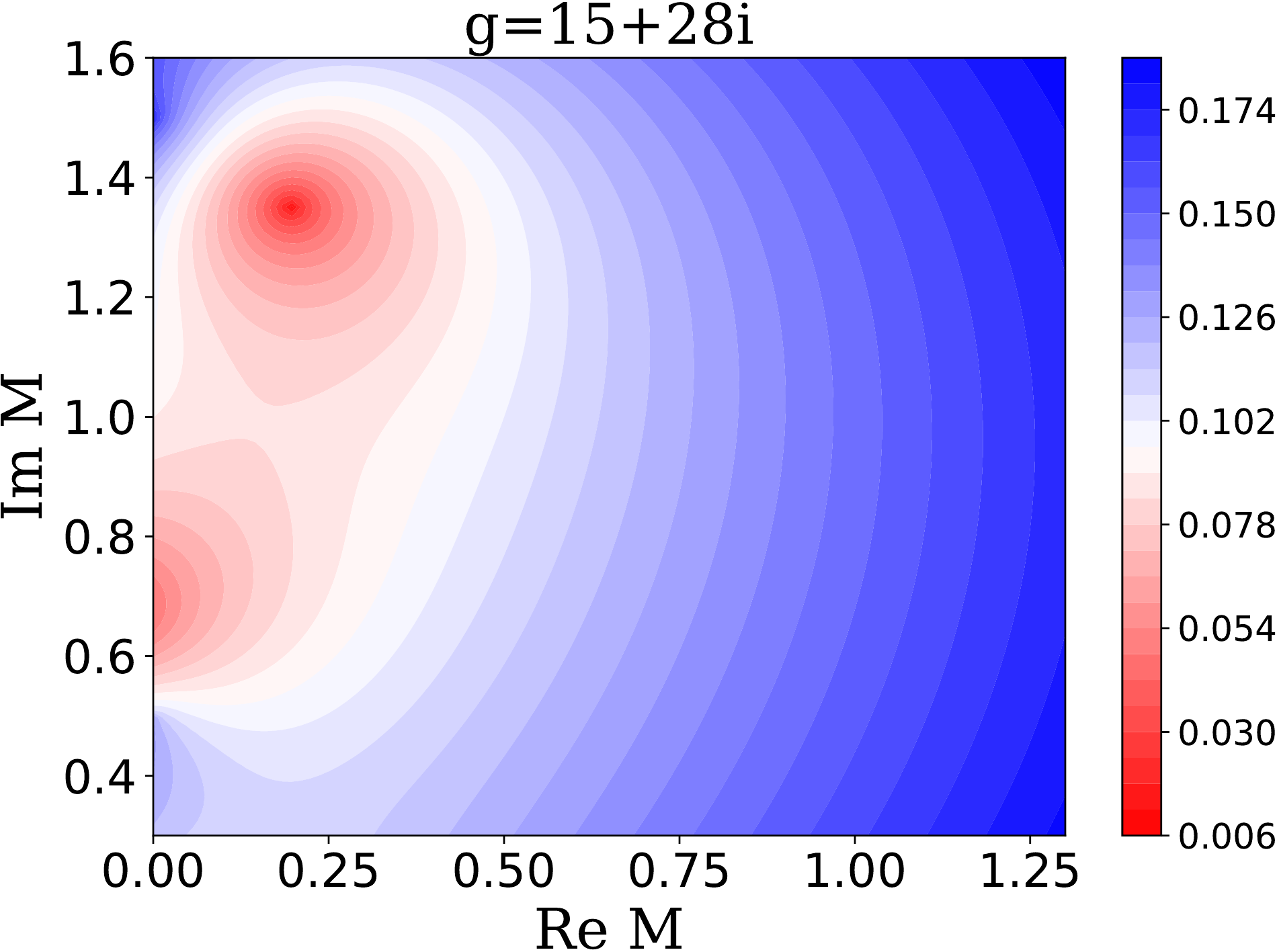}
	\put(-212,145){\Large (c)}
	\caption{\label{eq:sps}$\sqrt{|\der S/\der M|}$ on the complex $M$ plane at $\muu_5=0.5$ for three values of complex $g$.}
\end{figure}

By monitoring the magnitude of the gradient $\der S/\der M$ we found an interesting mechanism that changes the number of saddle points of $S$. In Figure~\ref{eq:sps}(a), (b) and (c) we display $\sqrt{|\der S/\der M|}$ as a function of $M$ for three values of $g$. (The square root of the gradient was taken for better visibility of the figures.) In Figure~\ref{eq:sps}(a) there is only one saddle point. (Of course there is another saddle for $M\to -M$.) When we increase the imaginary part of $g$, as shown in Figure~\ref{eq:sps}(b), a new saddle point is suddenly born out of the imaginary axis of $M$. So there are now two saddle points. When the imaginary part of $g$ is further increased, as shown in Figure~\ref{eq:sps}(c), the old saddle is absorbed into the imaginary axis and we are left with a single saddle. In this fashion the number of saddles (i.e., the solutions to \eqref{eq:hgdf8}) can jump abruptly. 

When there are multiple saddles, the dominant one is definitely the one that has the lowest value of $\Re\,S$, since it has the largest magnitude of $\rme^{-S}$ and dominates the  partition function as a sum over saddles $Z\approx\sum_n c_n \rme^{-S_n}$ provided that all $c_n$'s are of the same order. Following \cite{UedaPRL2019,Iskin2020}, we define three phases as below.
\begin{itemize}[noitemsep]
	\item Normal phase: \eqref{eq:hgdf8} has no solution, i.e., $M=0$ is the only saddle of $S$.
	\item Metastable \csb phase: there are solutions to \eqref{eq:hgdf8}, but their $\Re\;S$ are higher than that for $M=0$.
	\item Stable \csb phase: there are solutions to \eqref{eq:hgdf8} whose $\Re\;S$ are lower than that for $M=0$.
\end{itemize}
\begin{figure}[tb]
	\centering
	\includegraphics[width=.46\textwidth]{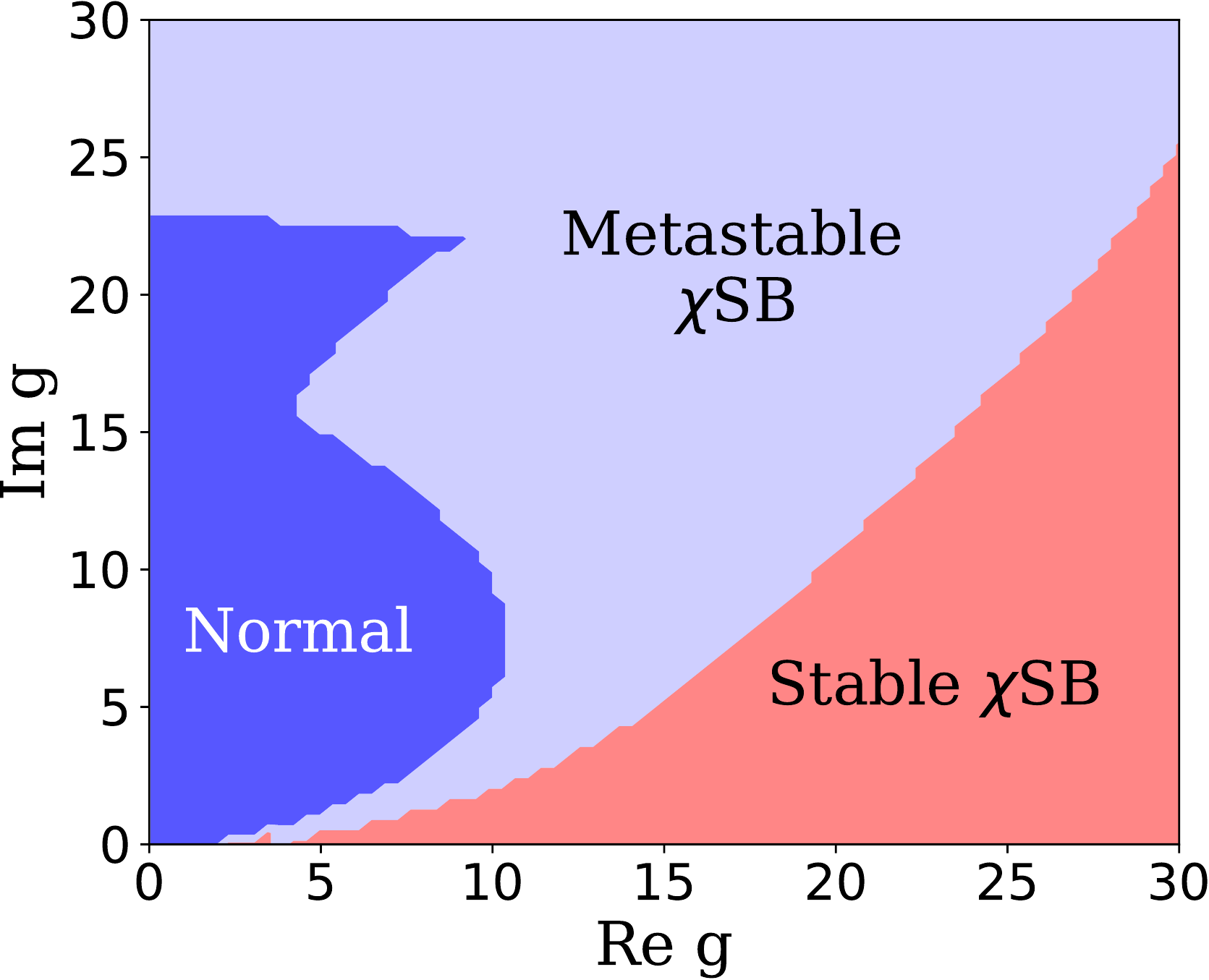}~
	\includegraphics[width=.49\textwidth]{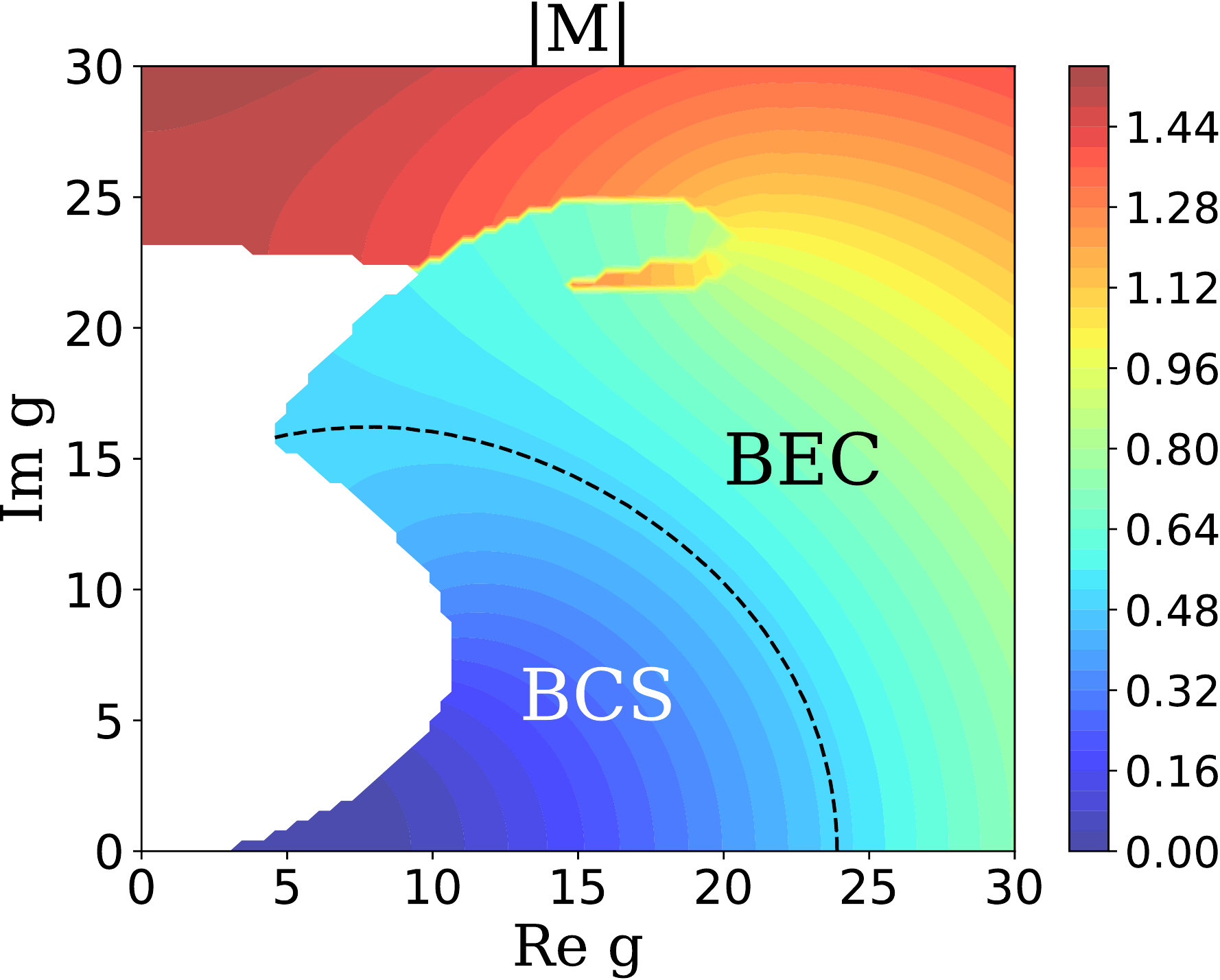}
	\caption{\label{eq:32425}The phase diagram for $\muu_5=0.5$ (left) and the magnitude of the complex dynamical mass $|M|$ (right). The dashed black line represents $|M|=\muu_5$.}
\end{figure}
In Figure~\ref{eq:32425} (left) we display the phase diagram for $\muu_5=0.5$ on the complex $g$ plane. In the vicinity of the real axis we have a stable \csb phase. As $\Im\;g$ increases, we are driven into a metastable \csb phase via a quantum phase transition. At small $\Re\;g$ the metastable saddle goes away and the chiral symmetry is completely restored. Interestingly, as one goes up along the imaginary axis, a metastable \csb state emerges suddenly at $\Im\;g\simeq 22.5$; this means that a very strong dissipation can trigger \csb (albeit a metastable one).  The global phase structure we found here is quite similar to the one for nonrelativistic fermions on a lattice \cite{UedaPRL2019}, though we note that the phase diagram at small $\Re\;g$ was not presented in \cite{UedaPRL2019} because of the limitation of numerical calculations. (See section~\ref{app} where we argue that part of the features of this phase diagram can be attributed to the usage of a sharp momentum cutoff.)

To gain more insights, in Figure~\ref{eq:32425} (right) we plot the magnitude of the gap $|M|$. When there are multiple saddles, we took the one that has the lowest $\Re\;S$. Notice that nothing dramatic happens at the boundary between the metastable \csb phase and the stable \csb phase. It is worth noting that the gap magnitude tends to be \emph{enhanced} by dissipation. As a crude guide we drew the boundary $|M|\sim \muu_5$ between the BCS regime and the BEC regime. We emphasize that this is a rule of thumb and a more rigorous characterization of the crossover region in non-Hermitian superfluids is left as an open problem. We point out that for $\Im\;g\gtrsim 20$ the gap reaches the UV cutoff scale ($|M|\sim 1$); this implies that all the calculations for $\Im\;g\gtrsim 20$ are extremely sensitive to the regularization scheme used, and hence one has to be careful about physical interpretations. 

\begin{figure}[tb]
	\centering
	\includegraphics[width=.49\textwidth]{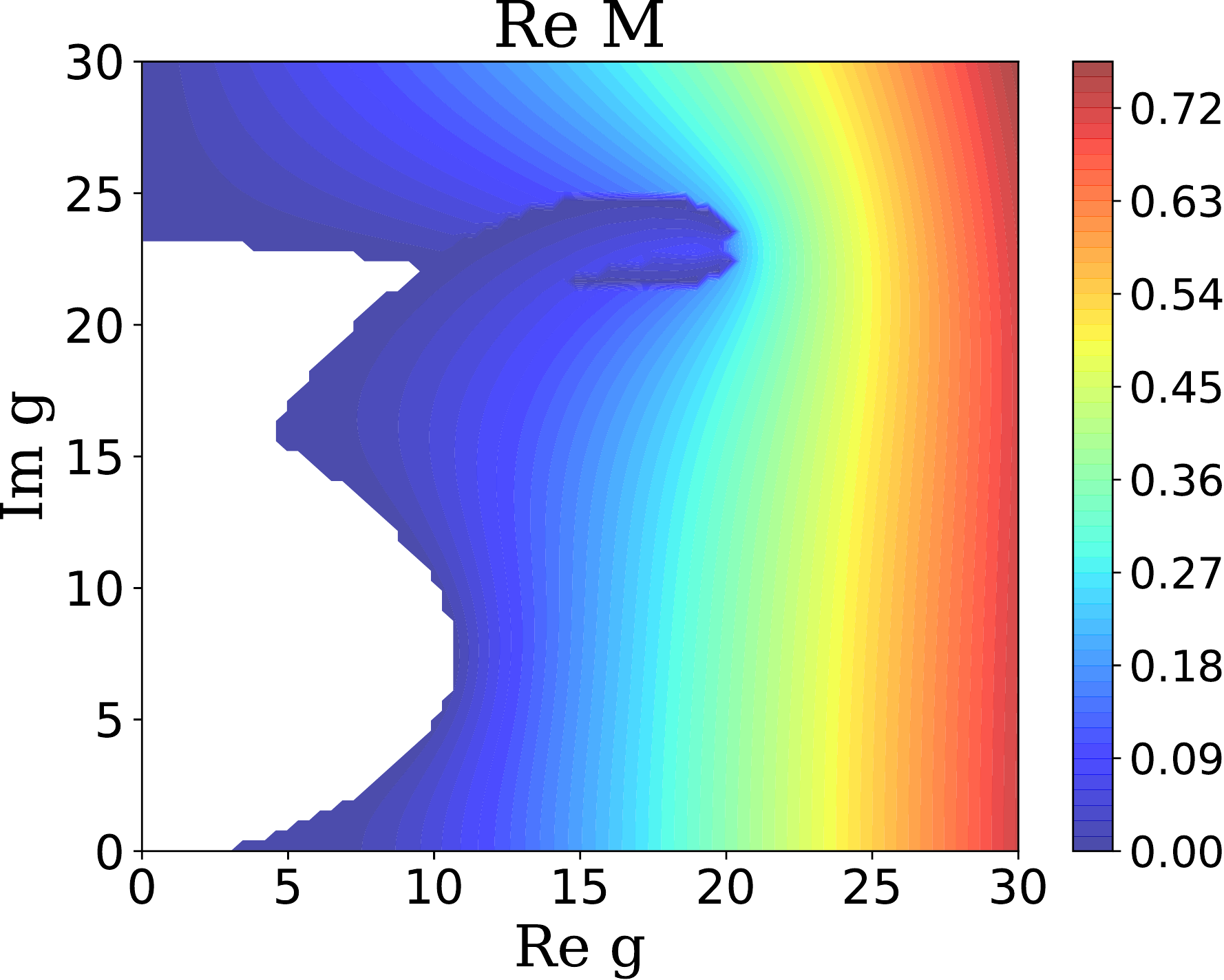}~
	\includegraphics[width=.49\textwidth]{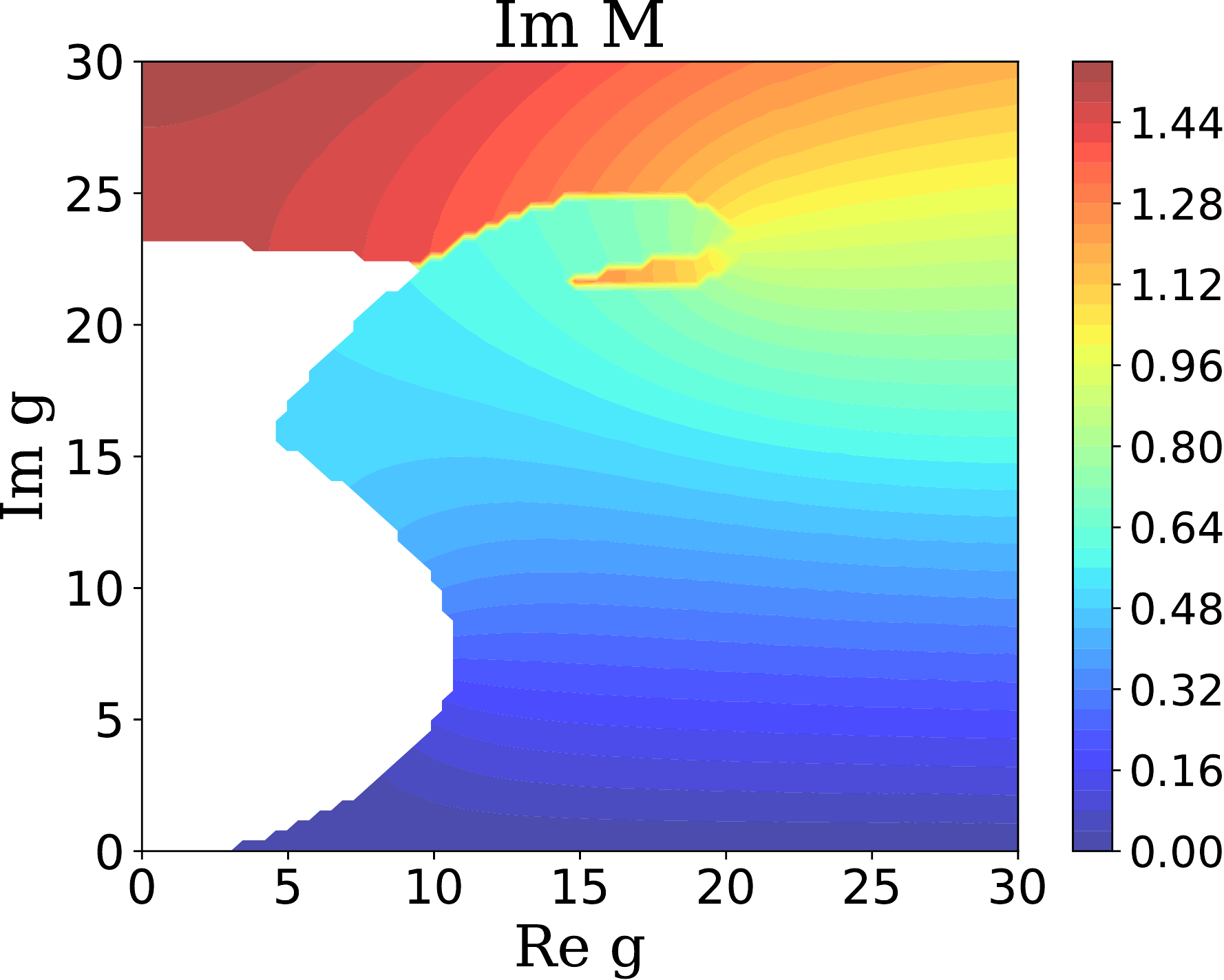}
	\caption{\label{eq:MM4561}The real and imaginary parts of the complex dynamical mass $M$ for $\muu_5=0.5$.}
\end{figure}

Figure~\ref{eq:MM4561} shows the real and imaginary parts of $M$. The left panel shows that $\Re\;M$ grows monotonically with $\Re\;g$ and is largely independent of $\Im\;g$. Note that $\Re\;M$ approaches zero along the boundary with the normal phase. This means that, when one moves out of the normal phase, a nontrivial solution to the gap equation emerges out of the imaginary axis of $M$.  The right panel shows that $\Im\;M$ grows monotonically with $\Im\;g$.

It is instructive to compare our findings with preceding works. 
\begin{itemize}[noitemsep]
	\item Ref.~\cite{UedaPRL2019} found an enhancement of superfluidity by dissipation, and attributed it to the continuous quantum Zeno effect which suppresses tunneling and reinforces on-site molecule formation. This effect is unique to a lattice system. Nevertheless we observed a similar enhancement in a continuum model; we speculate that the presence of a sharp momentum cutoff plays a role similar to that of a lattice. See section~\ref{app} for a discussion on an alternative UV regularization. 
	\item Ref.~\cite{Iskin2020} solved the BCS-BEC crossover for a complex scattering length and found that the superfluid phase is stable even in the limit of strong dissipation, which is surprising. The difference of \cite{Iskin2020} from \cite{UedaPRL2019} and this paper may be due to the fact that \cite{Iskin2020} considered a complex-valued chemical potential.%
	\footnote{A complex-valued chemical potential has been widely used in Monte Carlo simulations of QCD as a means of mitigating the sign problem \cite{deForcrand:2002hgr,DElia:2002tig}.}
	\item Ref.~\cite{Zhou2018} reports that a non-Hermitian perturbation (an imaginary magnetic field) enhances superfluidity.
	\item Ref.~\cite{Felski:2020vrm} solved the NJL model supplemented with a non-Hermitian bilinear term that preserves chiral symmetry. It was found that, as the non-Hermitian coupling grows, the dynamical mass first rises and then drops to zero.
\end{itemize}

\begin{figure}[tb]
	\centering
	\includegraphics[width=.49\textwidth]{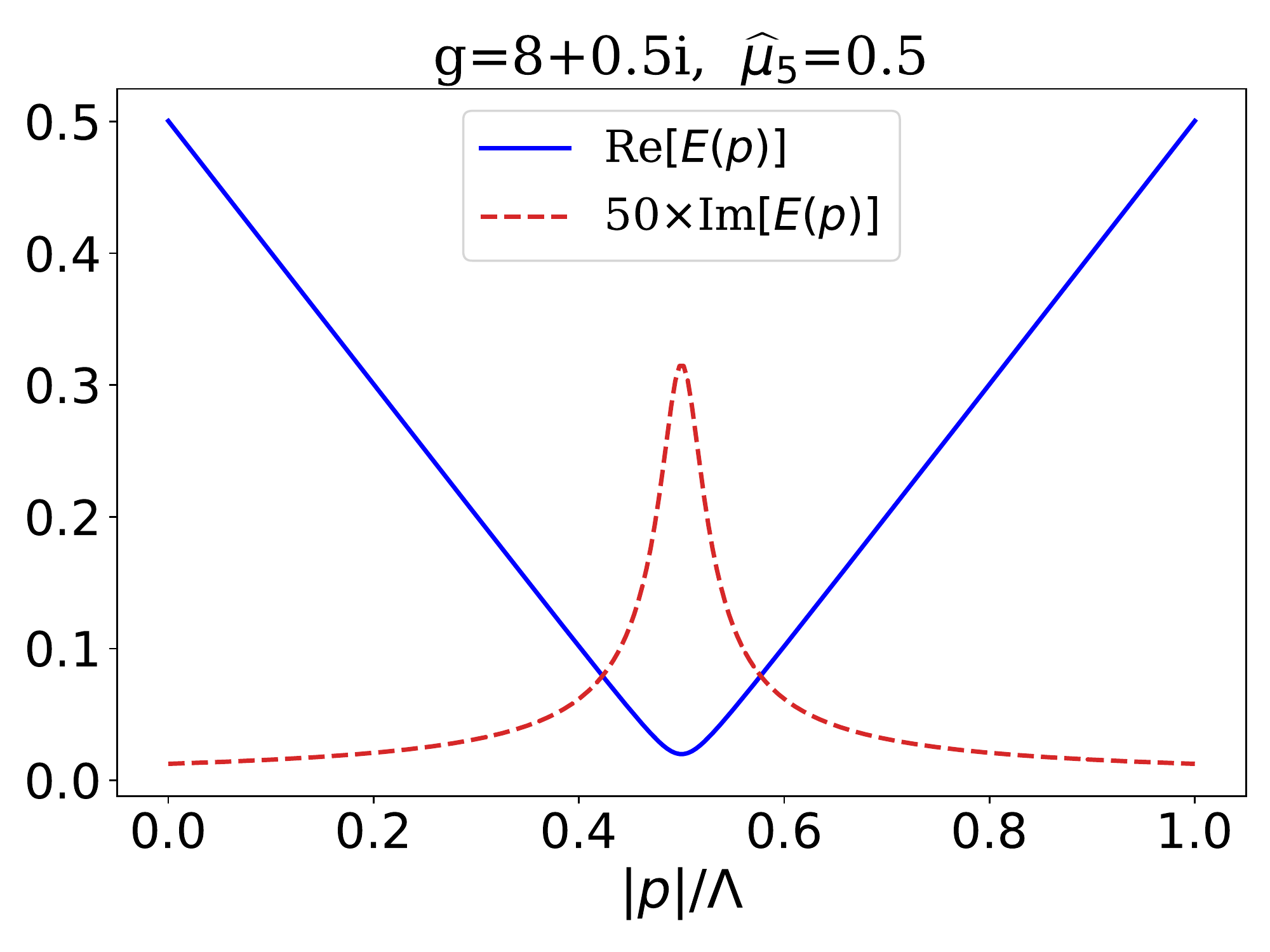}~
	\includegraphics[width=.49\textwidth]{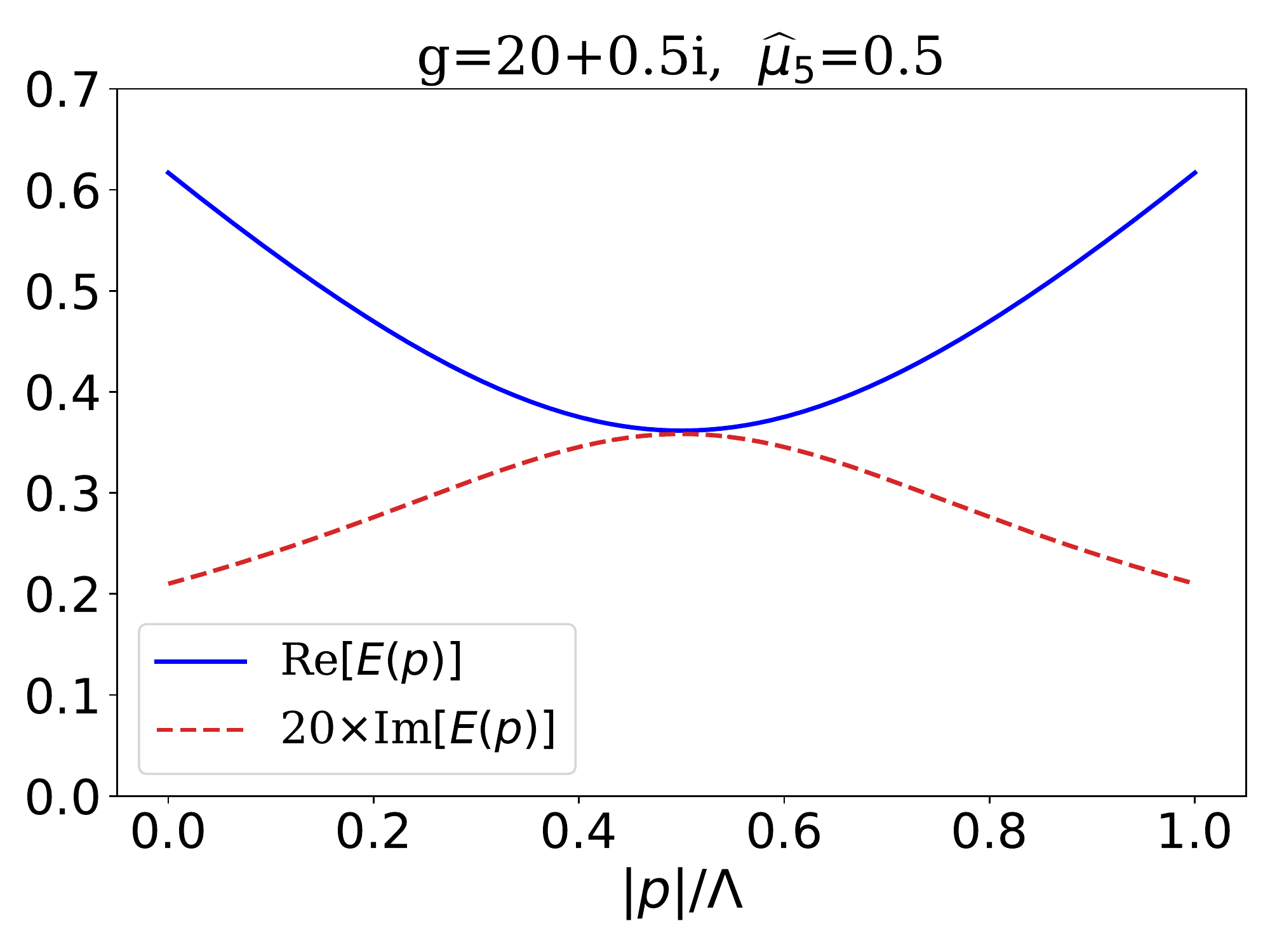}
	\caption{\label{eq:spec2384}The complex energy spectra of quasiparticles for weak (left) and strong (right) coupling.}
\end{figure}

Finally we turn to the quasiparticle excitation spectra $E(\pp)=\sqrt{(|\pp|/\Lambda-\muu_5)^2+M^2}$ which is generally complex for complex $g$. The real and (scaled) imaginary parts of $E(\pp)$ are displayed in Figure~\ref{eq:spec2384}. At weak coupling, the imaginary part of $E(\pp)$ is sharply concentrated around the Fermi level (left panel). By contrast, at stronger coupling, the imaginary part has a much broader support (right panel). These are results for weak dissipation ($\Im\;g=0.5$). If dissipation is stronger, a more peculiar thing can occur: at the boundary between the normal phase and the metastable \csb phase, $M$ is purely imaginary (cf.~Figure~\ref{eq:MM4561}), hence $E(\pp)$ is \emph{purely imaginary} for $\muu_5 - |M| \leq|\pp|/\Lambda\leq\muu_5 + |M|$ and is \emph{real} otherwise. The bounds $|\pp|/\Lambda=\muu_5 \pm |M|$ are an example of the so-called \emph{exceptional points} \cite{Heiss_2012} where the dimensionality of the eigenspace of the Hamiltonian decreases. 

\section{Dependence on the UV regularization scheme\label{app}}

In this section, we investigate what happens to the phase diagram when the sharp momentum cutoff used in section~\ref{sc:complex} is replaced with a more smooth UV regulator. Since the numerical search of a solution to the gap equation requires high numerical precision, it is mandatory to choose a regulator such that the momentum integral for the action $S$ can be evaluated analytically. For this reason, we choose smooth form factors 
\ba
	\frac{1}{\kkakko{\displaystyle 1+\mkakko{\frac{|\mathbf{p}|+\mu_5}{\Lambda}}^2}^3} \qquad \text{and}\qquad 
	\frac{1}{\kkakko{\displaystyle 1+\mkakko{\frac{|\mathbf{p}|-\mu_5}{\Lambda}}^2}^3}\,. 
	\label{eq:dshgsgfdfsd}
\ea
Then the action, in dimensionless form, reads as
\ba
	S & = \frac{M^2}{2g} - \frac{1}{2\pi^2}\int_0^{\infty}
	\rmd x\; x^2 \ckakko{
		\frac{\sqrt{(x+\muu_5)^2+M^2}}{[1+\mkakko{x+\muu_5}^2]^3} 
		+ 
		\frac{\sqrt{(x-\muu_5)^2+M^2}}{[1+\mkakko{x-\muu_5}^2]^3}
	}
	\\
	& = \frac{M^2}{2g} - \frac{1}{4\pi^2}\int_{-\infty}^{\infty}
	\rmd x\; x^2 \ckakko{
		\frac{\sqrt{(x+\muu_5)^2+M^2}}{[1+\mkakko{x+\muu_5}^2]^3} 
		+ 
		\frac{\sqrt{(x-\muu_5)^2+M^2}}{[1+\mkakko{x-\muu_5}^2]^3}
	}
	\\
	& = \frac{M^2}{2g} - \frac{1}{4\pi^2}\int_{-\infty}^{\infty}
	\rmd x\; [(x-\muu_5)^2+(x+\muu_5)^2] 
	\frac{\sqrt{x^2+M^2}}{(1+x^2)^3}
	\\
	& = \frac{M^2}{2g} - \frac{1}{2\pi^2}\int_{-\infty}^{\infty}
	\rmd x\; (x^2+\muu_5^2) 
	\frac{\sqrt{x^2+M^2}}{(1+x^2)^3}
	\\
	& = \frac{M^2}{2g} - \frac{1}{2\pi^2}
	\frac{\sqrt{M^2-1}\kkakko{M^2-2+(3M^2-2)\muu_5^2} + M^2 \kkakko{M^2+(3M^2-4) \muu_5^2 } \mathrm{ArcSec}\sqrt{M^2}}{4(M^2-1)^{3/2}}. 
\ea
Then we have for the gradient
\ba
\scalebox{0.85}{$\displaystyle 
	\frac{1}{2M}\frac{\der S}{\der M} = \frac{1}{2g} - 
	\frac{\sqrt{M^2-1}\kkakko{M^2+2+3(M^2-2)\muu_5^2} + \ckakko{M^2(M^2-4)+\kkakko{M^2(3M^2-8)+8}\muu_5^2}\mathrm{ArcSec}\sqrt{M^2}}{16\pi^2(M^2-1)^{5/2}}
	$}.
	\label{eq:254w}
\ea
\begin{figure}[tb]
	\centering
	\includegraphics[width=.46\textwidth]{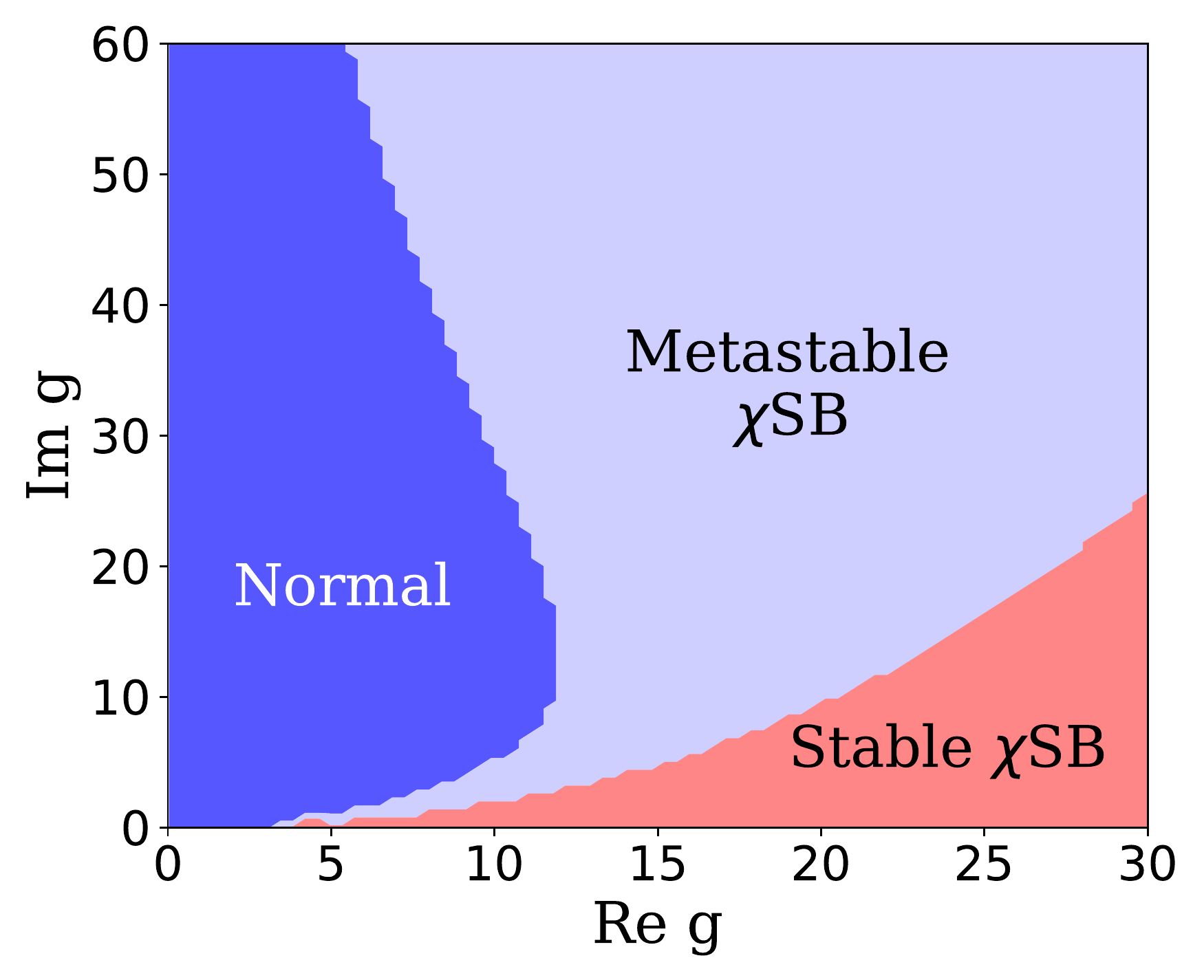}~
	\includegraphics[width=.49\textwidth]{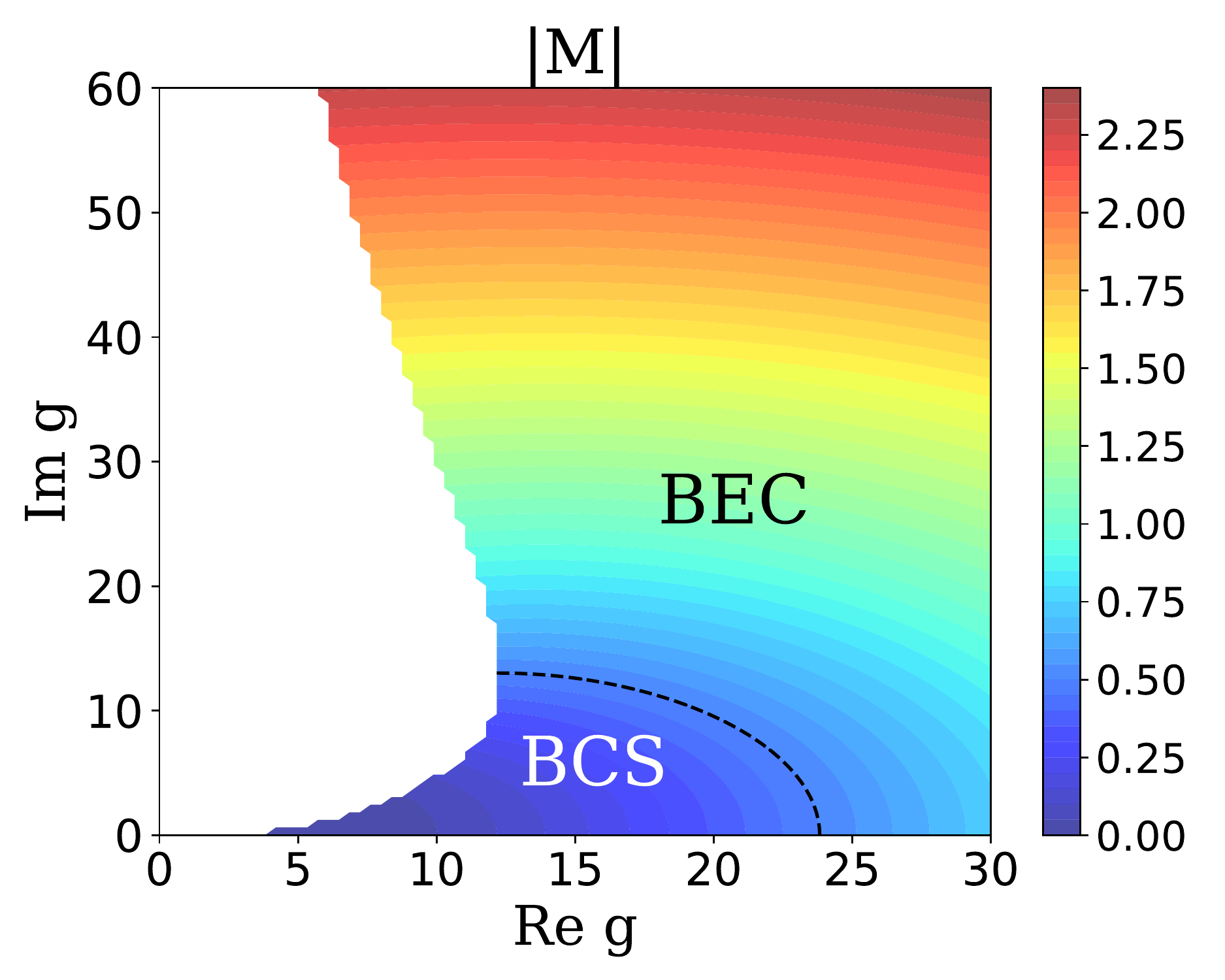}
	\caption{\label{eq:werf54}Same as Figure~\ref{eq:32425} but with a smooth UV regularization using the form factor \eqref{eq:dshgsgfdfsd}. Note that the vertical scale is twice larger than Figure~\ref{eq:32425}.}
\end{figure}%
We numerically searched for a solution to \eqref{eq:254w} with varying $g\in\CC$ and, if there was any solution, investigated its stability as compared to the trivial vacuum $M=0$. The result is shown in Figure~\ref{eq:werf54}. It is clear that the normal phase without a metastable vacuum has expanded substantially in comparison to Figure~\ref{eq:32425}. It is thus concluded that the emergence of a metastable \csb phase in the region $\text{Re}\;g\ll \text{Im}\; g$ in Figure~\ref{eq:32425} is largely an artifact due to the sharp momentum cutoff.

\section{\label{sc:conc}Conclusions and outlook}
In summary, we have investigated \csb of Dirac fermions in four dimensions at finite chiral chemical potential for a complex-valued four-fermion coupling. We varied the coupling over the entire range from the weakly bound BCS regime to the tightly bound BEC regime, and numerically constructed the phase diagram. Our primary result is that the imaginary coupling tends to enhance \csb up to a certain threshold; when the imaginary coupling exceeds the threshold, the chiral symmetry is restored, although a nontrivial solution to the gap equation continues to exist. We illustrated how complex saddles of the action come into existence and go away through the imaginary axis of the complex gap plane along which the action has a branch cut. We also worked out the complex energy spectra of quasiparticles. Our results can in principle be tested in experiments using ultracold atomic gases and other materials that host Dirac fermions \cite{Vafek:2013mpa,Wehling_2014,Armitage:2017cjs}.%
\footnote{It is conceivable that the creation of a chirally imbalanced matter in experiments is highly challenging. Creating a chirally balanced Dirac fermions at finite quark chemical potential would be easier, for which our results would be qualitatively correct if the dynamical mass is replaced with a superfluid gap.} However, it may be difficult to draw implications for quark matter in compact stars from this work, because the pointlike four-fermion interaction is a very crude approximation to the non-Abelian gauge interaction between quarks. 

There are miscellaneous future directions in which this work can be extended. A partial list is given below.
\begin{itemize}[noitemsep]
	\item To incorporate the competition between the chiral condensate and the diquark condensate (i.e., a competition between a Dirac mass and a Majorana mass) along the lines of \cite{Berges:1998rc,Ratti:2004ra}
	\item To investigate collective fluctuations around the saddle points and test the validity of the mean-field approximation for non-Hermitian \csb
	\item To analytically prove numerical findings in this work, e.g., the existence of a nontrivial solution to the gap equation for large $\Im\;g$
	\item To examine dissipative \csb under an external magnetic field that catalyzes \csb \cite{Andersen:2014xxa,Miransky:2015ava}
	\item To study $\U(1)_A$ vortices
	\item To study the effect of nonzero $\mu$ on non-Hermitian \csb
	\item To analyze the structure of Cooper pairs as a function of the complex coupling, and provide a more precise description of the non-Hermitian BCS-BEC crossover
	\item To use the renormalization group to improve the mean-field analysis
	\item To clarify how to apply the Lefschetz thimble approach \cite{Alexandru:2020wrj} to non-Hermitian \csb where the complex action and its gradient have branch cuts
	\item To use the results of this paper to benchmark algorithms (such as the complex Langevin method \cite{Aarts:2009uq}) for simulating complex-action theories
	\item To extend this work to lower spatial dimensions
	\item To test the conclusions of this paper with other low-energy effective models such as holography \cite{Herzog:2009xv} (see \cite{Arean:2019pom} for a non-Hermitian extension)
\end{itemize}

\bibliography{draft_v4.bbl}

\providecommand{\href}[2]{#2}\begingroup\raggedright\begin{thebibliography}{100}

\bibitem{Moiseyev2011}
N.~Moiseyev, \emph{{{Non-Hermitian Quantum Mechanics}}}, {{Cambridge University
  Press}} (2011),
  \href{https://doi.org/10.1017/CBO9780511976186}{10.1017/CBO9780511976186}.

\bibitem{Ashida:2020dkc}
Y.~Ashida, Z.~Gong and M.~Ueda, \emph{{{Non-Hermitian Physics}}},
  \href{https://arxiv.org/abs/2006.01837}{{\ttfamily 2006.01837}}.

\bibitem{Dalibard:1992zz}
J.~Dalibard, Y.~Castin and K.~Molmer, \emph{{Wave-function approach to
  dissipative processes in quantum optics}},
  \href{https://doi.org/10.1103/PhysRevLett.68.580}{\emph{Phys. Rev. Lett.}
  {\bfseries 68} (1992) 580}.

\bibitem{Carmichael_1993}
H.J.~Carmichael, \emph{Quantum trajectory theory for cascaded open systems},
  \href{https://doi.org/10.1103/PhysRevLett.70.2273}{\emph{Phys. Rev. Lett.}
  {\bfseries 70} (1993) 2273}.

\bibitem{Daley_2014}
A.J.~Daley, \emph{Quantum trajectories and open many-body quantum systems},
  \href{https://doi.org/10.1080/00018732.2014.933502}{\emph{Adv. Phys.}
  {\bfseries 63} (2014) 77} [\href{https://arxiv.org/abs/1405.6694}{{\ttfamily
  1405.6694}}].

\bibitem{Bender:1998ke}
C.M.~Bender and S.~Boettcher, \emph{{Real Spectra in Non-Hermitian Hamiltonians
  Having PT Symmetry}},
  \href{https://doi.org/10.1103/PhysRevLett.80.5243}{\emph{Phys. Rev. Lett.}
  {\bfseries 80} (1998) 5243}
  [\href{https://arxiv.org/abs/physics/9712001}{{\ttfamily physics/9712001}}].

\bibitem{Bender:2002vv}
C.M.~Bender, D.C.~Brody and H.F.~Jones, \emph{{Complex extension of quantum
  mechanics}}, \href{https://doi.org/10.1103/PhysRevLett.89.270401}{\emph{Phys.
  Rev. Lett.} {\bfseries 89} (2002) 270401}
  [\href{https://arxiv.org/abs/quant-ph/0208076}{{\ttfamily
  quant-ph/0208076}}].

\bibitem{Bender:2007nj}
C.M.~Bender, \emph{{Making sense of non-Hermitian Hamiltonians}},
  \href{https://doi.org/10.1088/0034-4885/70/6/R03}{\emph{Rept. Prog. Phys.}
  {\bfseries 70} (2007) 947}
  [\href{https://arxiv.org/abs/hep-th/0703096}{{\ttfamily hep-th/0703096}}].

\bibitem{Takasu2020pt}
Y.~Takasu, T.~Yagami, Y.~Ashida, R.~Hamazaki, Y.~Kuno and Y.~Takahashi,
  \emph{{PT-symmetric non-Hermitian quantum many-body system using ultracold
  atoms in an optical lattice with controlled dissipation}},
  \href{https://doi.org/10.1093/ptep/ptaa094}{\emph{Prog. Theor. Exp. Phys.}
  {\bfseries 2020} (2020) 12A110}
  [\href{https://arxiv.org/abs/2004.05734}{{\ttfamily 2004.05734}}].

\bibitem{bernard2001classification}
D.~Bernard and A.~LeClair, \emph{{{A Classification of Non-Hermitian Random
  Matrices}}},  in \emph{Statistical Field Theories. NATO Science Series
  (Series II: Mathematics, Physics and Chemistry)}, A.~Cappelli and
  G.~Mussardo, eds., vol.~73, pp.~207--214, Springer, Dordrecht, 2002,
  \href{https://doi.org/10.1007/978-94-010-0514-2_19}{DOI}
  [\href{https://arxiv.org/abs/cond-mat/0110649}{{\ttfamily
  cond-mat/0110649}}].

\bibitem{Kawabata:2018gjv}
K.~Kawabata, K.~Shiozaki, M.~Ueda and M.~Sato, \emph{{Symmetry and Topology in
  Non-Hermitian Physics}},
  \href{https://doi.org/10.1103/PhysRevX.9.041015}{\emph{Phys. Rev. X}
  {\bfseries 9} (2019) 041015}
  [\href{https://arxiv.org/abs/1812.09133}{{\ttfamily 1812.09133}}].

\bibitem{Martinez_Alvarez_PRB_2018}
V.M.~Martinez~Alvarez, J.E.~Barrios~Vargas and L.E.F.~Foa~Torres,
  \emph{{Non-Hermitian robust edge states in one dimension: Anomalous
  localization and eigenspace condensation at exceptional points}},
  \href{https://doi.org/10.1103/physrevb.97.121401}{\emph{Phys. Rev. B}
  {\bfseries 97} (2018) 121401}
  [\href{https://arxiv.org/abs/1711.05235}{{\ttfamily 1711.05235}}].

\bibitem{Yao_2018}
S.~Yao and Z.~Wang, \emph{{Edge States and Topological Invariants of
  Non-Hermitian Systems}},
  \href{https://doi.org/10.1103/physrevlett.121.086803}{\emph{Phys. Rev. Lett.}
  {\bfseries 121} (2018) 086803}
  [\href{https://arxiv.org/abs/1803.01876}{{\ttfamily 1803.01876}}].

\bibitem{Martinez_Alvarez_EPJ_2018}
V.M.~Martinez~Alvarez, J.E.~Barrios~Vargas, M.~Berdakin and L.E.F.~Foa~Torres,
  \emph{{Topological states of non-Hermitian systems}},
  \href{https://doi.org/10.1140/epjst/e2018-800091-5}{\emph{Eur. Phys. J.
  Special Topics} {\bfseries 227} (2018) 1295}
  [\href{https://arxiv.org/abs/1805.08200}{{\ttfamily 1805.08200}}].

\bibitem{Borgnia:2020mkg}
D.S.~Borgnia, A.J.~Kruchkov and R.-J.~Slager, \emph{{Non-Hermitian Boundary
  Modes and Topology}},
  \href{https://doi.org/10.1103/PhysRevLett.124.056802}{\emph{Phys. Rev. Lett.}
  {\bfseries 124} (2020) 056802}
  [\href{https://arxiv.org/abs/1902.07217}{{\ttfamily 1902.07217}}].

\bibitem{Okuma_2020}
N.~Okuma, K.~Kawabata, K.~Shiozaki and M.~Sato, \emph{{Topological Origin of
  Non-Hermitian Skin Effects}},
  \href{https://doi.org/10.1103/physrevlett.124.086801}{\emph{Phys. Rev. Lett.}
  {\bfseries 124} (2020) 086801}
  [\href{https://arxiv.org/abs/1910.02878}{{\ttfamily 1910.02878}}].

\bibitem{Hatano_1996}
N.~Hatano and D.R.~Nelson, \emph{{Localization Transitions in Non-Hermitian
  Quantum Mechanics}},
  \href{https://doi.org/10.1103/physrevlett.77.570}{\emph{Phys. Rev. Lett.}
  {\bfseries 77} (1996) 570}
  [\href{https://arxiv.org/abs/cond-mat/9603165}{{\ttfamily
  cond-mat/9603165}}].

\bibitem{Rothkopf:2011db}
A.~Rothkopf, T.~Hatsuda and S.~Sasaki, \emph{{Complex Heavy-Quark Potential at
  Finite Temperature from Lattice QCD}},
  \href{https://doi.org/10.1103/PhysRevLett.108.162001}{\emph{Phys. Rev. Lett.}
  {\bfseries 108} (2012) 162001}
  [\href{https://arxiv.org/abs/1108.1579}{{\ttfamily 1108.1579}}].

\bibitem{Rothkopf:2019ipj}
A.~Rothkopf, \emph{{Heavy Quarkonium in Extreme Conditions}},
  \href{https://doi.org/10.1016/j.physrep.2020.02.006}{\emph{Phys. Rept.}
  {\bfseries 858} (2020) 1} [\href{https://arxiv.org/abs/1912.02253}{{\ttfamily
  1912.02253}}].

\bibitem{Akamatsu:2020ypb}
Y.~Akamatsu, \emph{{Quarkonium in Quark-Gluon Plasma: Open Quantum System
  Approaches Re-examined}},  \href{https://arxiv.org/abs/2009.10559}{{\ttfamily
  2009.10559}}.

\bibitem{Chernodub:2019ggz}
M.N.~Chernodub and A.~Cortijo, \emph{{Non-Hermitian Chiral Magnetic Effect in
  Equilibrium}}, \href{https://doi.org/10.3390/sym12050761}{\emph{Symmetry}
  {\bfseries 12} (2020) 761}
  [\href{https://arxiv.org/abs/1901.06167}{{\ttfamily 1901.06167}}].

\bibitem{Denbleyker:2007dy}
A.~Denbleyker, D.~Du, Y.~Meurice and A.~Velytsky, \emph{{Fisher's Zeros and
  Perturbative Series in Gluodynamics}},
  \href{https://doi.org/10.22323/1.042.0269}{\emph{PoS} {\bfseries LATTICE2007}
  (2007) 269} [\href{https://arxiv.org/abs/0710.5771}{{\ttfamily 0710.5771}}].

\bibitem{Denbleyker:2010sv}
A.~Denbleyker, D.~Du, Y.~Liu, Y.~Meurice and H.~Zou, \emph{{Fisher's zeros as
  boundary of renormalization group flows in complex coupling spaces}},
  \href{https://doi.org/10.1103/PhysRevLett.104.251601}{\emph{Phys. Rev. Lett.}
  {\bfseries 104} (2010) 251601}
  [\href{https://arxiv.org/abs/1005.1993}{{\ttfamily 1005.1993}}].

\bibitem{Meurice:2011nf}
Y.~Meurice and H.~Zou, \emph{{Complex RG flows for 2D nonlinear O(N) sigma
  models}}, \href{https://doi.org/10.1103/PhysRevD.83.056009}{\emph{Phys. Rev.
  D} {\bfseries 83} (2011) 056009}
  [\href{https://arxiv.org/abs/1101.1319}{{\ttfamily 1101.1319}}].

\bibitem{Liu:2011zzh}
Y.~Liu and Y.~Meurice, \emph{{Lines of Fisher's zeros as separatrices for
  complex renormalization group flows}},
  \href{https://doi.org/10.1103/PhysRevD.83.096008}{\emph{Phys. Rev. D}
  {\bfseries 83} (2011) 096008}
  [\href{https://arxiv.org/abs/1103.4846}{{\ttfamily 1103.4846}}].

\bibitem{Denbleyker:2011aa}
A.~Denbleyker, A.~Bazavov, D.~Du, Y.~Liu, Y.~Meurice and H.~Zou,
  \emph{{Fisher's zeros, complex RG flows and confinement in LGT models}},
  \href{https://doi.org/10.22323/1.139.0299}{\emph{PoS} {\bfseries LATTICE2011}
  (2011) 299} [\href{https://arxiv.org/abs/1112.2734}{{\ttfamily 1112.2734}}].

\bibitem{Bazavov:2012ex}
A.~Bazavov, B.~Berg, D.~Du and Y.~Meurice, \emph{{Density of states and
  Fisher's zeros in compact U(1) pure gauge theory}},
  \href{https://doi.org/10.1103/PhysRevD.85.056010}{\emph{Phys. Rev. D}
  {\bfseries 85} (2012) 056010}
  [\href{https://arxiv.org/abs/1202.2109}{{\ttfamily 1202.2109}}].

\bibitem{Bender:2005hf}
C.M.~Bender, H.~Jones and R.~Rivers, \emph{{Dual PT-symmetric quantum field
  theories}}, \href{https://doi.org/10.1016/j.physletb.2005.08.087}{\emph{Phys.
  Lett. B} {\bfseries 625} (2005) 333}
  [\href{https://arxiv.org/abs/hep-th/0508105}{{\ttfamily hep-th/0508105}}].

\bibitem{Alexandre:2015kra}
J.~Alexandre, C.M.~Bender and P.~Millington, \emph{{Non-Hermitian extension of
  gauge theories and implications for neutrino physics}},
  \href{https://doi.org/10.1007/JHEP11(2015)111}{\emph{JHEP} {\bfseries 11}
  (2015) 111} [\href{https://arxiv.org/abs/1509.01203}{{\ttfamily
  1509.01203}}].

\bibitem{Alexandre:2017foi}
J.~Alexandre, P.~Millington and D.~Seynaeve, \emph{{Symmetries and conservation
  laws in non-Hermitian field theories}},
  \href{https://doi.org/10.1103/PhysRevD.96.065027}{\emph{Phys. Rev. D}
  {\bfseries 96} (2017) 065027}
  [\href{https://arxiv.org/abs/1707.01057}{{\ttfamily 1707.01057}}].

\bibitem{Beygi:2019qab}
A.~Beygi, S.~Klevansky and C.M.~Bender, \emph{{Relativistic $PT$-symmetric
  fermionic theories in 1+1 and 3+1 dimensions}},
  \href{https://doi.org/10.1103/PhysRevA.99.062117}{\emph{Phys. Rev. A}
  {\bfseries 99} (2019) 062117}
  [\href{https://arxiv.org/abs/1904.00878}{{\ttfamily 1904.00878}}].

\bibitem{Alexandre:2020wki}
J.~Alexandre, J.~Ellis and P.~Millington, \emph{{$\mathcal{PT}$ -symmetric
  non-Hermitian quantum field theories with supersymmetry}},
  \href{https://doi.org/10.1103/PhysRevD.101.085015}{\emph{Phys. Rev. D}
  {\bfseries 101} (2020) 085015}
  [\href{https://arxiv.org/abs/2001.11996}{{\ttfamily 2001.11996}}].

\bibitem{Alexandre:2020bet}
J.~Alexandre and N.E.~Mavromatos, \emph{{On the consistency of a non-Hermitian
  Yukawa interaction}},
  \href{https://doi.org/10.1016/j.physletb.2020.135562}{\emph{Phys. Lett. B}
  {\bfseries 807} (2020) 135562}
  [\href{https://arxiv.org/abs/2004.03699}{{\ttfamily 2004.03699}}].

\bibitem{Felski:2020vrm}
A.~Felski, A.~Beygi and S.~Klevansky, \emph{{Non-Hermitian extension of the
  Nambu--Jona-Lasinio model in 3+1 and 1+1 dimensions}},
  \href{https://doi.org/10.1103/PhysRevD.101.116001}{\emph{Phys. Rev. D}
  {\bfseries 101} (2020) 116001}
  [\href{https://arxiv.org/abs/2004.04011}{{\ttfamily 2004.04011}}].

\bibitem{Alexandre:2020tba}
J.~Alexandre, N.E.~Mavromatos and A.~Soto, \emph{{Dynamical Majorana Neutrino
  Masses and Axions}},
  \href{https://doi.org/10.1016/j.nuclphysb.2020.115212}{\emph{Nucl. Phys. B}
  {\bfseries 961} (2020) 115212}
  [\href{https://arxiv.org/abs/2004.04611}{{\ttfamily 2004.04611}}].

\bibitem{Mavromatos:2020hfy}
N.E.~Mavromatos and A.~Soto, \emph{{Dynamical Majorana Neutrino Masses and
  Axions II: Inclusion of Axial Background and Anomaly Terms}},
  \href{https://doi.org/10.1016/j.nuclphysb.2020.115275}{\emph{Nucl. Phys. B}
  {\bfseries 962} (2021) 115275}
  [\href{https://arxiv.org/abs/2006.13616}{{\ttfamily 2006.13616}}].

\bibitem{Chernodub:2020cew}
M.~Chernodub, A.~Cortijo and M.~Ruggieri, \emph{{Spontaneous Non-Hermiticity in
  Nambu-Jona-Lasinio model}},
  \href{https://arxiv.org/abs/2008.11629}{{\ttfamily 2008.11629}}.

\bibitem{Nambu:1961tp}
Y.~Nambu and G.~Jona-Lasinio, \emph{{Dynamical Model of Elementary Particles
  Based on an Analogy with Superconductivity. I}},
  \href{https://doi.org/10.1103/PhysRev.122.345}{\emph{Phys. Rev.} {\bfseries
  122} (1961) 345}.

\bibitem{Nambu:1961fr}
Y.~Nambu and G.~Jona-Lasinio, \emph{{Dynamical Model of Elementary Particles
  Based on an Analogy with Superconductivity. II}},
  \href{https://doi.org/10.1103/PhysRev.124.246}{\emph{Phys. Rev.} {\bfseries
  124} (1961) 246}.

\bibitem{Klevansky:1992qe}
S.~Klevansky, \emph{{The Nambu-Jona-Lasinio model of quantum chromodynamics}},
  \href{https://doi.org/10.1103/RevModPhys.64.649}{\emph{Rev. Mod. Phys.}
  {\bfseries 64} (1992) 649}.

\bibitem{Hatsuda:1994pi}
T.~Hatsuda and T.~Kunihiro, \emph{{QCD phenomenology based on a chiral
  effective Lagrangian}},
  \href{https://doi.org/10.1016/0370-1573(94)90022-1}{\emph{Phys. Rept.}
  {\bfseries 247} (1994) 221}
  [\href{https://arxiv.org/abs/hep-ph/9401310}{{\ttfamily hep-ph/9401310}}].

\bibitem{Buballa:2003qv}
M.~Buballa, \emph{{NJL model analysis of quark matter at large density}},
  \href{https://doi.org/10.1016/j.physrep.2004.11.004}{\emph{Phys. Rept.}
  {\bfseries 407} (2005) 205}
  [\href{https://arxiv.org/abs/hep-ph/0402234}{{\ttfamily hep-ph/0402234}}].

\bibitem{Guralnik:2007rx}
G.~Guralnik and Z.~Guralnik, \emph{{Complexified path integrals and the phases
  of quantum field theory}},
  \href{https://doi.org/10.1016/j.aop.2010.06.001}{\emph{Annals Phys.}
  {\bfseries 325} (2010) 2486}
  [\href{https://arxiv.org/abs/0710.1256}{{\ttfamily 0710.1256}}].

\bibitem{Alexanian:2008kd}
G.~Alexanian, R.~MacKenzie, M.~Paranjape and J.~Ruel, \emph{{Path integration
  and perturbation theory with complex Euclidean actions}},
  \href{https://doi.org/10.1103/PhysRevD.77.105014}{\emph{Phys. Rev. D}
  {\bfseries 77} (2008) 105014}
  [\href{https://arxiv.org/abs/0802.0354}{{\ttfamily 0802.0354}}].

\bibitem{Witten:2010cx}
E.~Witten, \emph{{Analytic Continuation Of Chern-Simons Theory}}, {\emph{AMS/IP
  Stud. Adv. Math.} {\bfseries 50} (2011) 347}
  [\href{https://arxiv.org/abs/1001.2933}{{\ttfamily 1001.2933}}].

\bibitem{Cristoforetti:2012su}
{\scshape AuroraScience} collaboration, \emph{{New approach to the sign problem
  in quantum field theories: High density QCD on a Lefschetz thimble}},
  \href{https://doi.org/10.1103/PhysRevD.86.074506}{\emph{Phys. Rev. D}
  {\bfseries 86} (2012) 074506}
  [\href{https://arxiv.org/abs/1205.3996}{{\ttfamily 1205.3996}}].

\bibitem{Fujii:2013sra}
H.~Fujii, D.~Honda, M.~Kato, Y.~Kikukawa, S.~Komatsu and T.~Sano, \emph{{Hybrid
  Monte Carlo on Lefschetz thimbles - A study of the residual sign problem}},
  \href{https://doi.org/10.1007/JHEP10(2013)147}{\emph{JHEP} {\bfseries 10}
  (2013) 147} [\href{https://arxiv.org/abs/1309.4371}{{\ttfamily 1309.4371}}].

\bibitem{Tanizaki:2014xba}
Y.~Tanizaki and T.~Koike, \emph{{Real-time Feynman path integral with
  Picard--Lefschetz theory and its applications to quantum tunneling}},
  \href{https://doi.org/10.1016/j.aop.2014.09.003}{\emph{Annals Phys.}
  {\bfseries 351} (2014) 250}
  [\href{https://arxiv.org/abs/1406.2386}{{\ttfamily 1406.2386}}].

\bibitem{Kanazawa:2014qma}
T.~Kanazawa and Y.~Tanizaki, \emph{{Structure of Lefschetz thimbles in simple
  fermionic systems}},
  \href{https://doi.org/10.1007/JHEP03(2015)044}{\emph{JHEP} {\bfseries 03}
  (2015) 044} [\href{https://arxiv.org/abs/1412.2802}{{\ttfamily 1412.2802}}].

\bibitem{Tanizaki:2015rda}
Y.~Tanizaki, Y.~Hidaka and T.~Hayata, \emph{{Lefschetz-thimble analysis of the
  sign problem in one-site fermion model}},
  \href{https://doi.org/10.1088/1367-2630/18/3/033002}{\emph{New J. Phys.}
  {\bfseries 18} (2016) 033002}
  [\href{https://arxiv.org/abs/1509.07146}{{\ttfamily 1509.07146}}].

\bibitem{Fujii:2015vha}
H.~Fujii, S.~Kamata and Y.~Kikukawa, \emph{{Monte Carlo study of Lefschetz
  thimble structure in one-dimensional Thirring model at finite density}},
  \href{https://doi.org/10.1007/JHEP12(2015)125}{\emph{JHEP} {\bfseries 12}
  (2015) 125} [\href{https://arxiv.org/abs/1509.09141}{{\ttfamily
  1509.09141}}].

\bibitem{Alexandru:2015sua}
A.~Alexandru, G.~Basar, P.F.~Bedaque, G.W.~Ridgway and N.C.~Warrington,
  \emph{{Sign problem and Monte Carlo calculations beyond Lefschetz thimbles}},
  \href{https://doi.org/10.1007/JHEP05(2016)053}{\emph{JHEP} {\bfseries 05}
  (2016) 053} [\href{https://arxiv.org/abs/1512.08764}{{\ttfamily
  1512.08764}}].

\bibitem{Alexandru:2015xva}
A.~Alexandru, G.~Basar and P.~Bedaque, \emph{{Monte Carlo algorithm for
  simulating fermions on Lefschetz thimbles}},
  \href{https://doi.org/10.1103/PhysRevD.93.014504}{\emph{Phys. Rev. D}
  {\bfseries 93} (2016) 014504}
  [\href{https://arxiv.org/abs/1510.03258}{{\ttfamily 1510.03258}}].

\bibitem{Mori:2017pne}
Y.~Mori, K.~Kashiwa and A.~Ohnishi, \emph{{Toward solving the sign problem with
  path optimization method}},
  \href{https://doi.org/10.1103/PhysRevD.96.111501}{\emph{Phys. Rev. D}
  {\bfseries 96} (2017) 111501}
  [\href{https://arxiv.org/abs/1705.05605}{{\ttfamily 1705.05605}}].

\bibitem{Alexandru:2017czx}
A.~Alexandru, P.F.~Bedaque, H.~Lamm and S.~Lawrence, \emph{{Deep Learning
  Beyond Lefschetz Thimbles}},
  \href{https://doi.org/10.1103/PhysRevD.96.094505}{\emph{Phys. Rev. D}
  {\bfseries 96} (2017) 094505}
  [\href{https://arxiv.org/abs/1709.01971}{{\ttfamily 1709.01971}}].

\bibitem{Fukuma:2019wbv}
M.~Fukuma, N.~Matsumoto and N.~Umeda, \emph{{Applying the tempered Lefschetz
  thimble method to the Hubbard model away from half-filling}},
  \href{https://doi.org/10.1103/PhysRevD.100.114510}{\emph{Phys. Rev. D}
  {\bfseries 100} (2019) 114510}
  [\href{https://arxiv.org/abs/1906.04243}{{\ttfamily 1906.04243}}].

\bibitem{Mou:2019gyl}
Z.-G.~Mou, P.M.~Saffin and A.~Tranberg, \emph{{Quantum tunnelling, real-time
  dynamics and Picard-Lefschetz thimbles}},
  \href{https://doi.org/10.1007/JHEP11(2019)135}{\emph{JHEP} {\bfseries 11}
  (2019) 135} [\href{https://arxiv.org/abs/1909.02488}{{\ttfamily
  1909.02488}}].

\bibitem{Alexandru:2020wrj}
A.~Alexandru, G.~Basar, P.F.~Bedaque and N.C.~Warrington, \emph{{Complex Paths
  Around The Sign Problem}},
  \href{https://arxiv.org/abs/2007.05436}{{\ttfamily 2007.05436}}.

\bibitem{Markum:1999yr}
H.~Markum, R.~Pullirsch and T.~Wettig, \emph{{NonHermitian random matrix theory
  and lattice QCD with chemical potential}},
  \href{https://doi.org/10.1103/PhysRevLett.83.484}{\emph{Phys. Rev. Lett.}
  {\bfseries 83} (1999) 484}
  [\href{https://arxiv.org/abs/hep-lat/9906020}{{\ttfamily hep-lat/9906020}}].

\bibitem{Khoruzhenko2009}
B.A.~Khoruzhenko and H.J.~Sommers, \emph{{{Non-Hermitian Random Matrix
  Ensembles}}},  \href{https://arxiv.org/abs/0911.5645}{{\ttfamily 0911.5645}}.

\bibitem{Kanazawabook}
T.~Kanazawa, \emph{{Dirac Spectra in Dense QCD}}, Springer Japan (2013).

\bibitem{LeggettBook2006}
A.J.~Leggett, \emph{{Quantum Liquids: Bose condensation and Cooper pairing in
  condensed-matter systems}}, Oxford University Press (2006).

\bibitem{Casalbuoni:2018haw}
R.~Casalbuoni, \emph{{Lecture Notes on Superconductivity: Condensed Matter and
  QCD}},  \href{https://arxiv.org/abs/1810.11125}{{\ttfamily 1810.11125}}.

\bibitem{Chtchelkatchev2012}
N.M.~Chtchelkatchev, A.A.~Golubov, T.I.~Baturina and V.M.~Vinokur,
  \emph{{Stimulation of the Fluctuation Superconductivity by PT Symmetry}},
  \href{https://doi.org/10.1103/PhysRevLett.109.150405}{\emph{Phys. Rev. Lett.}
  {\bfseries 109} (2012) 150405}
  [\href{https://arxiv.org/abs/arXiv:1008.3590}{{\ttfamily arXiv:1008.3590}}].

\bibitem{Ghatak2018}
A.~Ghatak and T.~Das, \emph{{Theory of superconductivity with non-Hermitian and
  parity-time reversal symmetric Cooper pairing symmetry}},
  \href{https://doi.org/10.1103/PhysRevB.97.014512}{\emph{Phys. Rev. B}
  {\bfseries 97} (2018) 014512}
  [\href{https://arxiv.org/abs/arXiv:1708.09108}{{\ttfamily
  arXiv:1708.09108}}].

\bibitem{Zhou2018}
L.~Zhou and X.~Cui, \emph{{Enhanced fermion pairing and superfluidity by an
  imaginary magnetic field}},
  \href{https://doi.org/10.1016/j.isci.2019.03.031}{\emph{{iScience}}
  {\bfseries 14} (2019) 257}
  [\href{https://arxiv.org/abs/1812.11008}{{\ttfamily 1812.11008}}].

\bibitem{UedaPRL2019}
K.~Yamamoto, M.~Nakagawa, K.~Adachi, K.~Takasan, M.~Ueda and N.~Kawakami,
  \emph{{Theory of Non-Hermitian Fermionic Superfluidity with a Complex-Valued
  Interaction}},
  \href{https://doi.org/10.1103/PhysRevLett.123.123601}{\emph{Phys. Rev. Lett.}
  {\bfseries 123} (2019) 123601}
  [\href{https://arxiv.org/abs/arXiv:1903.04720}{{\ttfamily
  arXiv:1903.04720}}].

\bibitem{Iskin2020}
M.~Iskin, \emph{{Non-Hermitian BCS-BEC evolution with a complex scattering
  length}}, \href{https://doi.org/10.1103/PhysRevA.103.013724}{\emph{Phys. Rev.
  A} {\bfseries 103} (2020) 013724}
  [\href{https://arxiv.org/abs/2002.00653}{{\ttfamily 2002.00653}}].

\bibitem{Blatt1962}
J.M.~Blatt, K.W.~B\"oer and W.~Brandt, \emph{{Bose-Einstein Condensation of
  Excitons}}, \href{https://doi.org/10.1103/PhysRev.126.1691}{\emph{Phys. Rev.}
  {\bfseries 126} (1962) 1691}.

\bibitem{Yoshioka_2011}
K.~Yoshioka, E.~Chae and M.~Kuwata-Gonokami, \emph{{Transition to a
  Bose-Einstein condensate and relaxation explosion of excitons at sub-Kelvin
  temperatures}}, \href{https://doi.org/10.1038/ncomms1335}{\emph{Nature
  Communications} {\bfseries 2} (2011) 328}
  [\href{https://arxiv.org/abs/1008.2431}{{\ttfamily 1008.2431}}].

\bibitem{Ogawa2007}
T.~Ogawa, Y.~Tomio and K.~Asano, \emph{{Quantum condensation in electron-hole
  systems: excitonic BEC-BCS crossover and biexciton crystallization}},
  \href{https://doi.org/10.1088/0953-8984/19/29/295205}{\emph{J. Phys.:
  Condens. Matter} {\bfseries 19} (2007) 295205}.

\bibitem{Yamaguchi2012}
M.~Yamaguchi, K.~Kamide, T.~Ogawa and Y.~Yamamoto, \emph{{BEC-BCS-laser
  crossover in Coulomb-correlated electron-hole-photon systems}},
  \href{https://doi.org/10.1088/1367-2630/14/6/065001}{\emph{New J. Phys.}
  {\bfseries 14} (2012) 065001}.

\bibitem{Hanai_JLTP_2016}
R.~Hanai, P.B.~Littlewood and Y.~Ohashi, \emph{{Non-equilibrium Properties of a
  Pumped-Decaying Bose-Condensed Electron-Hole Gas in the BCS-BEC Crossover
  Region}}, \href{https://doi.org/10.1007/s10909-016-1552-6}{\emph{J. Low Temp.
  Phys.} {\bfseries 183} (2016) 127}
  [\href{https://arxiv.org/abs/1506.08983}{{\ttfamily 1506.08983}}].

\bibitem{Hanai_PRB_2017}
R.~Hanai, P.B.~Littlewood and Y.~Ohashi, \emph{{Dynamical instability of a
  driven-dissipative electron-hole condensate in the BCS-BEC crossover
  region}}, \href{https://doi.org/10.1103/PhysRevB.96.125206}{\emph{Phys. Rev.
  B} {\bfseries 96} (2017) 125206}
  [\href{https://arxiv.org/abs/1610.08622}{{\ttfamily 1610.08622}}].

\bibitem{Kawamura_JLTP_2019}
T.~Kawamura, D.~Kagamihara, R.~Hanai and Y.~Ohashi, \emph{{Strong-Coupling
  Theory for a Non-equilibrium Unitary Fermi Gas}},
  \href{https://doi.org/10.1007/s10909-019-02310-7}{\emph{J. Low Temp. Phys.}
  (2019) }.

\bibitem{Kawamura_PRA_2020}
T.~Kawamura, R.~Hanai, D.~Kagamihara, D.~Inotani and Y.~Ohashi,
  \emph{{Nonequilibrium strong-coupling theory for a driven-dissipative
  ultracold Fermi gas in the BCS-BEC crossover region}},
  \href{https://doi.org/10.1103/PhysRevA.101.013602}{\emph{Phys. Rev. A}
  {\bfseries 101} (2020) 013602}
  [\href{https://arxiv.org/abs/1910.12476}{{\ttfamily 1910.12476}}].

\bibitem{Triola_2017}
C.~Triola, A.~Pertsova, R.S.~Markiewicz and A.V.~Balatsky, \emph{{{Excitonic
  gap formation in pumped Dirac materials}}},
  \href{https://doi.org/10.1103/physrevb.95.205410}{\emph{Phys. Rev. B}
  {\bfseries 95} (2017) 205410}
  [\href{https://arxiv.org/abs/1701.04206}{{\ttfamily 1701.04206}}].

\bibitem{Pertsova_2018}
A.~Pertsova and A.V.~Balatsky, \emph{{Excitonic instability in optically pumped
  three-dimensional Dirac materials}},
  \href{https://doi.org/10.1103/physrevb.97.075109}{\emph{Phys. Rev. B}
  {\bfseries 97} (2018) 075109}
  [\href{https://arxiv.org/abs/1710.09132}{{\ttfamily 1710.09132}}].

\bibitem{Pertsova_review2020}
A.~Pertsova and A.V.~Balatsky, \emph{{Dynamically Induced Excitonic Instability
  in Pumped Dirac Materials}},
  \href{https://doi.org/10.1002/andp.201900549}{\emph{Annalen der Physik}
  {\bfseries 532} (2020) 1900549}
  [\href{https://arxiv.org/abs/1912.09400}{{\ttfamily 1912.09400}}].

\bibitem{Nozieres1985}
P.~Nozi\`{e}res and S.~Schmitt-Rink, \emph{{Bose condensation in an attractive
  fermion gas: From weak to strong coupling superconductivity}},
  {\emph{{Journal of Low Temperature Physics}} {\bfseries 59} (1985) 195}.

\bibitem{Chen2005}
Q.~Chen, J.~Stajic, S.~Tan and K.~Levin, \emph{{BCS-BEC crossover: From high
  temperature superconductors to ultracold superfluids}},
  \href{https://doi.org/https://doi.org/10.1016/j.physrep.2005.02.005}{\emph{Physics
  Reports} {\bfseries 412} (2005) 1}
  [\href{https://arxiv.org/abs/arXiv:cond-mat/0404274}{{\ttfamily
  arXiv:cond-mat/0404274}}].

\bibitem{GiorginiRMP2008}
S.~Giorgini, L.P.~Pitaevskii and S.~Stringari, \emph{{Theory of ultracold
  atomic Fermi gases}},
  \href{https://doi.org/10.1103/RevModPhys.80.1215}{\emph{Rev. Mod. Phys.}
  {\bfseries 80} (2008) 1215}
  [\href{https://arxiv.org/abs/arXiv:0706.3360}{{\ttfamily arXiv:0706.3360}}].

\bibitem{ZwergerBook2012}
W.~Zwerger, ed., \emph{{The BCS-BEC Crossover and the Unitary Fermi Gas}},
  Springer (2012),
  \href{https://doi.org/10.1007/978-3-642-21978-8}{10.1007/978-3-642-21978-8}.

\bibitem{RanderiaBook2014}
M.~Randeria and E.~Taylor, \emph{{Crossover from Bardeen-Cooper-Schrieffer to
  Bose-Einstein Condensation and the Unitary Fermi Gas}},
  \href{https://doi.org/10.1146/annurev-conmatphys-031113-133829}{\emph{{Annual
  Review of Condensed Matter Physics}} {\bfseries 5} (2014) 209}
  [\href{https://arxiv.org/abs/arXiv:1306.5785}{{\ttfamily arXiv:1306.5785}}].

\bibitem{Strinati2018}
G.C.~Strinati, P.~Pieri, G.~R\"{o}pke, P.~Schuck and M.~Urban, \emph{{The
  BCS-BEC crossover: From ultra-cold Fermi gases to nuclear systems}},
  \href{https://doi.org/https://doi.org/10.1016/j.physrep.2018.02.004}{\emph{Physics
  Reports} {\bfseries 738} (2018) 1}
  [\href{https://arxiv.org/abs/arXiv:1802.05997}{{\ttfamily
  arXiv:1802.05997}}].

\bibitem{Zhu_2007}
S.-L.~Zhu, B.~Wang and L.-M.~Duan, \emph{{Simulation and Detection of Dirac
  Fermions with Cold Atoms in an Optical Lattice}},
  \href{https://doi.org/10.1103/physrevlett.98.260402}{\emph{Phys. Rev. Lett.}
  {\bfseries 98} (2007) 260402}
  [\href{https://arxiv.org/abs/cond-mat/0703454}{{\ttfamily
  cond-mat/0703454}}].

\bibitem{Lim_2009}
L.-K.~Lim, A.~Lazarides, A.~Hemmerich and C.~Morais~Smith, \emph{{Strongly
  interacting two-dimensional Dirac fermions}},
  \href{https://doi.org/10.1209/0295-5075/88/36001}{\emph{Europhys. Lett.}
  {\bfseries 88} (2009) 36001}
  [\href{https://arxiv.org/abs/0905.1281}{{\ttfamily 0905.1281}}].

\bibitem{IgnacioCirac:2010us}
J.~Ignacio~Cirac, P.~Maraner and J.K.~Pachos, \emph{{Cold atom simulation of
  interacting relativistic quantum field theories}},
  \href{https://doi.org/10.1103/PhysRevLett.105.190403}{\emph{Phys. Rev. Lett.}
  {\bfseries 105} (2010) 190403}
  [\href{https://arxiv.org/abs/1006.2975}{{\ttfamily 1006.2975}}].

\bibitem{Nishida:2005ds}
Y.~Nishida and H.~Abuki, \emph{{BCS-BEC crossover in a relativistic superfluid
  and its significance to quark matter}},
  \href{https://doi.org/10.1103/PhysRevD.72.096004}{\emph{Phys. Rev. D}
  {\bfseries 72} (2005) 096004}
  [\href{https://arxiv.org/abs/hep-ph/0504083}{{\ttfamily hep-ph/0504083}}].

\bibitem{Abuki:2006dv}
H.~Abuki, \emph{{BCS/BEC crossover in Quark Matter and Evolution of its Static
  and Dynamic properties: From the atomic unitary gas to color
  superconductivity}},
  \href{https://doi.org/10.1016/j.nuclphysa.2007.03.134}{\emph{Nucl. Phys. A}
  {\bfseries 791} (2007) 117}
  [\href{https://arxiv.org/abs/hep-ph/0605081}{{\ttfamily hep-ph/0605081}}].

\bibitem{Deng:2006ed}
J.~Deng, A.~Schmitt and Q.~Wang, \emph{{Relativistic BCS-BEC crossover in a
  boson-fermion model}},
  \href{https://doi.org/10.1103/PhysRevD.76.034013}{\emph{Phys. Rev. D}
  {\bfseries 76} (2007) 034013}
  [\href{https://arxiv.org/abs/nucl-th/0611097}{{\ttfamily nucl-th/0611097}}].

\bibitem{He:2007kd}
L.~He and P.~Zhuang, \emph{{Relativistic BCS-BEC crossover at zero
  temperature}}, \href{https://doi.org/10.1103/PhysRevD.75.096003}{\emph{Phys.
  Rev. D} {\bfseries 75} (2007) 096003}
  [\href{https://arxiv.org/abs/hep-ph/0703042}{{\ttfamily hep-ph/0703042}}].

\bibitem{He:2007yj}
L.~He and P.~Zhuang, \emph{{Relativistic BCS-BEC crossover at finite
  temperature and Its application to color superconductivity}},
  \href{https://doi.org/10.1103/PhysRevD.76.056003}{\emph{Phys. Rev. D}
  {\bfseries 76} (2007) 056003}
  [\href{https://arxiv.org/abs/0705.1634}{{\ttfamily 0705.1634}}].

\bibitem{Sun:2007fc}
G.-f.~Sun, L.~He and P.~Zhuang, \emph{{BEC-BCS crossover in the
  Nambu-Jona-Lasinio model of QCD}},
  \href{https://doi.org/10.1103/PhysRevD.75.096004}{\emph{Phys. Rev. D}
  {\bfseries 75} (2007) 096004}
  [\href{https://arxiv.org/abs/hep-ph/0703159}{{\ttfamily hep-ph/0703159}}].

\bibitem{Brauner:2008td}
T.~Brauner, \emph{{BCS-BEC crossover in dense relativistic matter: Collective
  excitations}}, \href{https://doi.org/10.1103/PhysRevD.77.096006}{\emph{Phys.
  Rev. D} {\bfseries 77} (2008) 096006}
  [\href{https://arxiv.org/abs/0803.2422}{{\ttfamily 0803.2422}}].

\bibitem{Chatterjee:2008dr}
B.~Chatterjee, H.~Mishra and A.~Mishra, \emph{{BCS-BEC crossover and phase
  structure of relativistic systems: A variational approach}},
  \href{https://doi.org/10.1103/PhysRevD.79.014003}{\emph{Phys. Rev. D}
  {\bfseries 79} (2009) 014003}
  [\href{https://arxiv.org/abs/0804.1051}{{\ttfamily 0804.1051}}].

\bibitem{He:2010nb}
L.~He, \emph{{Nambu-Jona-Lasinio model description of weakly interacting Bose
  condensate and BEC-BCS crossover in dense QCD-like theories}},
  \href{https://doi.org/10.1103/PhysRevD.82.096003}{\emph{Phys. Rev. D}
  {\bfseries 82} (2010) 096003}
  [\href{https://arxiv.org/abs/1007.1920}{{\ttfamily 1007.1920}}].

\bibitem{Kanazawa:2011tt}
T.~Kanazawa, T.~Wettig and N.~Yamamoto, \emph{{Singular values of the Dirac
  operator in dense QCD-like theories}},
  \href{https://doi.org/10.1007/JHEP12(2011)007}{\emph{JHEP} {\bfseries 12}
  (2011) 007} [\href{https://arxiv.org/abs/1110.5858}{{\ttfamily 1110.5858}}].

\bibitem{He:2013gga}
L.~He, S.~Mao and P.~Zhuang, \emph{{BCS-BEC crossover in relativistic Fermi
  systems}}, \href{https://doi.org/10.1142/S0217751X13300548}{\emph{Int. J.
  Mod. Phys. A} {\bfseries 28} (2013) 1330054}
  [\href{https://arxiv.org/abs/1311.6704}{{\ttfamily 1311.6704}}].

\bibitem{Alford:2007xm}
M.G.~Alford, A.~Schmitt, K.~Rajagopal and T.~Sch\"{a}fer, \emph{{Color
  superconductivity in dense quark matter}},
  \href{https://doi.org/10.1103/RevModPhys.80.1455}{\emph{Rev. Mod. Phys.}
  {\bfseries 80} (2008) 1455}
  [\href{https://arxiv.org/abs/0709.4635}{{\ttfamily 0709.4635}}].

\bibitem{Fukushima:2008xe}
K.~Fukushima, D.E.~Kharzeev and H.J.~Warringa, \emph{{The Chiral Magnetic
  Effect}}, \href{https://doi.org/10.1103/PhysRevD.78.074033}{\emph{Phys. Rev.
  D} {\bfseries 78} (2008) 074033}
  [\href{https://arxiv.org/abs/0808.3382}{{\ttfamily 0808.3382}}].

\bibitem{Yamamoto:2011gk}
A.~Yamamoto, \emph{{Chiral magnetic effect in lattice QCD with a chiral
  chemical potential}},
  \href{https://doi.org/10.1103/PhysRevLett.107.031601}{\emph{Phys. Rev. Lett.}
  {\bfseries 107} (2011) 031601}
  [\href{https://arxiv.org/abs/1105.0385}{{\ttfamily 1105.0385}}].

\bibitem{Ruggieri:2011xc}
M.~Ruggieri, \emph{{The Critical End Point of Quantum Chromodynamics Detected
  by Chirally Imbalanced Quark Matter}},
  \href{https://doi.org/10.1103/PhysRevD.84.014011}{\emph{Phys. Rev. D}
  {\bfseries 84} (2011) 014011}
  [\href{https://arxiv.org/abs/1103.6186}{{\ttfamily 1103.6186}}].

\bibitem{Gatto:2011wc}
R.~Gatto and M.~Ruggieri, \emph{{Hot Quark Matter with an Axial Chemical
  Potential}}, \href{https://doi.org/10.1103/PhysRevD.85.054013}{\emph{Phys.
  Rev. D} {\bfseries 85} (2012) 054013}
  [\href{https://arxiv.org/abs/1110.4904}{{\ttfamily 1110.4904}}].

\bibitem{Braguta:2015zta}
V.~Braguta, V.~Goy, E.M.~Ilgenfritz, A.Y.~Kotov, A.~Molochkov,
  M.~Muller-Preussker et~al., \emph{{Two-Color QCD with Non-zero Chiral
  Chemical Potential}},
  \href{https://doi.org/10.1007/JHEP06(2015)094}{\emph{JHEP} {\bfseries 06}
  (2015) 094} [\href{https://arxiv.org/abs/1503.06670}{{\ttfamily
  1503.06670}}].

\bibitem{Xu:2015vna}
S.-S.~Xu, Z.-F.~Cui, B.~Wang, Y.-M.~Shi, Y.-C.~Yang and H.-S.~Zong,
  \emph{{Chiral phase transition with a chiral chemical potential in the
  framework of Dyson-Schwinger equations}},
  \href{https://doi.org/10.1103/PhysRevD.91.056003}{\emph{Phys. Rev. D}
  {\bfseries 91} (2015) 056003}
  [\href{https://arxiv.org/abs/1505.00316}{{\ttfamily 1505.00316}}].

\bibitem{Braguta:2016aov}
V.~Braguta and A.Y.~Kotov, \emph{{Catalysis of Dynamical Chiral Symmetry
  Breaking by Chiral Chemical Potential}},
  \href{https://doi.org/10.1103/PhysRevD.93.105025}{\emph{Phys. Rev. D}
  {\bfseries 93} (2016) 105025}
  [\href{https://arxiv.org/abs/1601.04957}{{\ttfamily 1601.04957}}].

\bibitem{Braguta:2019pxt}
V.~Braguta, M.~Katsnelson, A.~Kotov and A.~Trunin, \emph{{Catalysis of
  Dynamical Chiral Symmetry Breaking by Chiral Chemical Potential in Dirac
  semimetals}}, \href{https://doi.org/10.1103/PhysRevB.100.085117}{\emph{Phys.
  Rev. B} {\bfseries 100} (2019) 085117}
  [\href{https://arxiv.org/abs/1904.07003}{{\ttfamily 1904.07003}}].

\bibitem{Redlich:1984md}
A.~Redlich and L.~Wijewardhana, \emph{{Induced Chern-simons Terms at High
  Temperatures and Finite Densities}},
  \href{https://doi.org/10.1103/PhysRevLett.54.970}{\emph{Phys. Rev. Lett.}
  {\bfseries 54} (1985) 970}.

\bibitem{Rubakov:1985nk}
V.~Rubakov, \emph{{On the Electroweak Theory at High Fermion Density}},
  \href{https://doi.org/10.1143/PTP.75.366}{\emph{Prog. Theor. Phys.}
  {\bfseries 75} (1986) 366}.

\bibitem{Akamatsu:2013pjd}
Y.~Akamatsu and N.~Yamamoto, \emph{{Chiral Plasma Instabilities}},
  \href{https://doi.org/10.1103/PhysRevLett.111.052002}{\emph{Phys. Rev. Lett.}
  {\bfseries 111} (2013) 052002}
  [\href{https://arxiv.org/abs/1302.2125}{{\ttfamily 1302.2125}}].

\bibitem{dlmf_cosh}
{{NIST Digital Library of Mathematical Functions}},
  \emph{\rm\url{https://dlmf.nist.gov/4.36\#E2}}, .

\bibitem{Abuki:2001be}
H.~Abuki, T.~Hatsuda and K.~Itakura, \emph{{Structural change of Cooper pairs
  and momentum dependent gap in color superconductivity}},
  \href{https://doi.org/10.1103/PhysRevD.65.074014}{\emph{Phys. Rev. D}
  {\bfseries 65} (2002) 074014}
  [\href{https://arxiv.org/abs/hep-ph/0109013}{{\ttfamily hep-ph/0109013}}].

\bibitem{Itakura:2002vr}
K.~Itakura, \emph{{Structure change of Cooper pairs in color superconductivity:
  Crossover from BCS to BEC?}},
  \href{https://doi.org/10.1016/S0375-9474(02)01523-3}{\emph{Nucl. Phys. A}
  {\bfseries 715} (2003) 859}
  [\href{https://arxiv.org/abs/hep-ph/0209081}{{\ttfamily hep-ph/0209081}}].

\bibitem{Chandrasekhar1962}
B.S.~Chandrasekhar, \emph{{A NOTE ON THE MAXIMUM CRITICAL FIELD OF HIGH-FIELD
  SUPERCONDUCTORS}}, \href{https://doi.org/10.1063/1.1777362}{\emph{Appl. Phys.
  Lett.} {\bfseries 1} (1962) 7}.

\bibitem{Clogston1962}
A.M.~Clogston, \emph{{Upper Limit for the Critical Field in Hard
  Superconductors}},
  \href{https://doi.org/10.1103/PhysRevLett.9.266}{\emph{Phys. Rev. Lett.}
  {\bfseries 9} (1962) 266}.

\bibitem{Radzihovsky_2010}
L.~Radzihovsky and D.E.~Sheehy, \emph{{Imbalanced Feshbach-resonant Fermi
  gases}}, \href{https://doi.org/10.1088/0034-4885/73/7/076501}{\emph{Rep.
  Prog. Phys.} {\bfseries 73} (2010) 076501}.

\bibitem{Chevy_2010}
F.~Chevy and C.~Mora, \emph{{Ultra-cold polarized Fermi gases}},
  \href{https://doi.org/10.1088/0034-4885/73/11/112401}{\emph{Rep. Prog. Phys.}
  {\bfseries 73} (2010) 112401}.

\bibitem{Alford:2000ze}
M.G.~Alford, J.A.~Bowers and K.~Rajagopal, \emph{{Crystalline color
  superconductivity}},
  \href{https://doi.org/10.1103/PhysRevD.63.074016}{\emph{Phys. Rev. D}
  {\bfseries 63} (2001) 074016}
  [\href{https://arxiv.org/abs/hep-ph/0008208}{{\ttfamily hep-ph/0008208}}].

\bibitem{Splittorff:2000mm}
K.~Splittorff, D.~Son and M.A.~Stephanov, \emph{{QCD - like theories at finite
  baryon and isospin density}},
  \href{https://doi.org/10.1103/PhysRevD.64.016003}{\emph{Phys. Rev. D}
  {\bfseries 64} (2001) 016003}
  [\href{https://arxiv.org/abs/hep-ph/0012274}{{\ttfamily hep-ph/0012274}}].

\bibitem{Bedaque:2001je}
P.F.~Bedaque and T.~Sch\"{a}fer, \emph{{High density quark matter under
  stress}}, \href{https://doi.org/10.1016/S0375-9474(01)01272-6}{\emph{Nucl.
  Phys. A} {\bfseries 697} (2002) 802}
  [\href{https://arxiv.org/abs/hep-ph/0105150}{{\ttfamily hep-ph/0105150}}].

\bibitem{Casalbuoni:2003wh}
R.~Casalbuoni and G.~Nardulli, \emph{{Inhomogeneous superconductivity in
  condensed matter and QCD}},
  \href{https://doi.org/10.1103/RevModPhys.76.263}{\emph{Rev. Mod. Phys.}
  {\bfseries 76} (2004) 263}
  [\href{https://arxiv.org/abs/hep-ph/0305069}{{\ttfamily hep-ph/0305069}}].

\bibitem{Nickel:2009wj}
D.~Nickel, \emph{{Inhomogeneous phases in the Nambu-Jona-Lasino and quark-meson
  model}}, \href{https://doi.org/10.1103/PhysRevD.80.074025}{\emph{Phys. Rev.
  D} {\bfseries 80} (2009) 074025}
  [\href{https://arxiv.org/abs/0906.5295}{{\ttfamily 0906.5295}}].

\bibitem{Buballa:2014tba}
M.~Buballa and S.~Carignano, \emph{{Inhomogeneous chiral condensates}},
  \href{https://doi.org/10.1016/j.ppnp.2014.11.001}{\emph{Prog. Part. Nucl.
  Phys.} {\bfseries 81} (2015) 39}
  [\href{https://arxiv.org/abs/1406.1367}{{\ttfamily 1406.1367}}].

\bibitem{deForcrand:2002hgr}
P.~de~Forcrand and O.~Philipsen, \emph{{The QCD phase diagram for small
  densities from imaginary chemical potential}},
  \href{https://doi.org/10.1016/S0550-3213(02)00626-0}{\emph{Nucl. Phys. B}
  {\bfseries 642} (2002) 290}
  [\href{https://arxiv.org/abs/hep-lat/0205016}{{\ttfamily hep-lat/0205016}}].

\bibitem{DElia:2002tig}
M.~D'Elia and M.-P.~Lombardo, \emph{{Finite density QCD via imaginary chemical
  potential}}, \href{https://doi.org/10.1103/PhysRevD.67.014505}{\emph{Phys.
  Rev. D} {\bfseries 67} (2003) 014505}
  [\href{https://arxiv.org/abs/hep-lat/0209146}{{\ttfamily hep-lat/0209146}}].

\bibitem{Heiss_2012}
W.D.~Heiss, \emph{The physics of exceptional points},
  \href{https://doi.org/10.1088/1751-8113/45/44/444016}{\emph{J. Phys. A: Math.
  Theor.} {\bfseries 45} (2012) 444016}
  [\href{https://arxiv.org/abs/1210.7536}{{\ttfamily 1210.7536}}].

\bibitem{Vafek:2013mpa}
O.~Vafek and A.~Vishwanath, \emph{{Dirac Fermions in Solids: From High-Tc
  cuprates and Graphene to Topological Insulators and Weyl Semimetals}},
  \href{https://doi.org/10.1146/annurev-conmatphys-031113-133841}{\emph{Ann.
  Rev. Condensed Matter Phys.} {\bfseries 5} (2014) 83}
  [\href{https://arxiv.org/abs/1306.2272}{{\ttfamily 1306.2272}}].

\bibitem{Wehling_2014}
T.~Wehling, A.~Black-Schaffer and A.~Balatsky, \emph{Dirac materials},
  \href{https://doi.org/10.1080/00018732.2014.927109}{\emph{Adv. Phys.}
  {\bfseries 63} (2014) 1} [\href{https://arxiv.org/abs/1405.5774}{{\ttfamily
  1405.5774}}].

\bibitem{Armitage:2017cjs}
N.~Armitage, E.~Mele and A.~Vishwanath, \emph{{Weyl and Dirac Semimetals in
  Three Dimensional Solids}},
  \href{https://doi.org/10.1103/RevModPhys.90.015001}{\emph{Rev. Mod. Phys.}
  {\bfseries 90} (2018) 015001}
  [\href{https://arxiv.org/abs/1705.01111}{{\ttfamily 1705.01111}}].

\bibitem{Berges:1998rc}
J.~Berges and K.~Rajagopal, \emph{{Color superconductivity and chiral symmetry
  restoration at nonzero baryon density and temperature}},
  \href{https://doi.org/10.1016/S0550-3213(98)00620-8}{\emph{Nucl. Phys. B}
  {\bfseries 538} (1999) 215}
  [\href{https://arxiv.org/abs/hep-ph/9804233}{{\ttfamily hep-ph/9804233}}].

\bibitem{Ratti:2004ra}
C.~Ratti and W.~Weise, \emph{{Thermodynamics of two-colour QCD and the Nambu
  Jona-Lasinio model}},
  \href{https://doi.org/10.1103/PhysRevD.70.054013}{\emph{Phys. Rev. D}
  {\bfseries 70} (2004) 054013}
  [\href{https://arxiv.org/abs/hep-ph/0406159}{{\ttfamily hep-ph/0406159}}].

\bibitem{Andersen:2014xxa}
J.O.~Andersen, W.R.~Naylor and A.~Tranberg, \emph{{Phase diagram of QCD in a
  magnetic field: A review}},
  \href{https://doi.org/10.1103/RevModPhys.88.025001}{\emph{Rev. Mod. Phys.}
  {\bfseries 88} (2016) 025001}
  [\href{https://arxiv.org/abs/1411.7176}{{\ttfamily 1411.7176}}].

\bibitem{Miransky:2015ava}
V.A.~Miransky and I.A.~Shovkovy, \emph{{Quantum field theory in a magnetic
  field: From quantum chromodynamics to graphene and Dirac semimetals}},
  \href{https://doi.org/10.1016/j.physrep.2015.02.003}{\emph{Phys. Rept.}
  {\bfseries 576} (2015) 1} [\href{https://arxiv.org/abs/1503.00732}{{\ttfamily
  1503.00732}}].

\bibitem{Aarts:2009uq}
G.~Aarts, E.~Seiler and I.-O.~Stamatescu, \emph{{The Complex Langevin method:
  When can it be trusted?}},
  \href{https://doi.org/10.1103/PhysRevD.81.054508}{\emph{Phys. Rev. D}
  {\bfseries 81} (2010) 054508}
  [\href{https://arxiv.org/abs/0912.3360}{{\ttfamily 0912.3360}}].

\bibitem{Herzog:2009xv}
C.P.~Herzog, \emph{{Lectures on Holographic Superfluidity and
  Superconductivity}},
  \href{https://doi.org/10.1088/1751-8113/42/34/343001}{\emph{J. Phys. A}
  {\bfseries 42} (2009) 343001}
  [\href{https://arxiv.org/abs/0904.1975}{{\ttfamily 0904.1975}}].

\bibitem{Arean:2019pom}
D.~Are\'an, K.~Landsteiner and I.~Salazar~Landea, \emph{{Non-Hermitian
  Holography}},
  \href{https://doi.org/10.21468/SciPostPhys.9.3.032}{\emph{SciPost Phys.}
  {\bfseries 9} (2020) 032} [\href{https://arxiv.org/abs/1912.06647}{{\ttfamily
  1912.06647}}].

\end{thebibliography}\endgroup
\end{document}